\documentclass[10pt, conference]{IEEEtran}
\usepackage{amsmath, amssymb, amsthm}
\usepackage{mathtools}
\usepackage{xfrac}
\usepackage[table]{xcolor}
\usepackage{tabularx}
\usepackage[inline]{enumitem}
\usepackage[hidelinks]{hyperref}
\usepackage{calc}
\usepackage{balance}
\usepackage{cite}
\usepackage{tcolorbox}
	
\usepackage{graphicx}
\usepackage[export]{adjustbox}

\usepackage[T1]{fontenc}
\usepackage[scaled=0.85]{beramono}

\usepackage{xcolor}
\usepackage{tikz,pgfplots}
\usepackage{ifthen}

\makeatletter
\def\hlinewd#1{%
\noalign{\ifnum0=`}\fi\hrule \@height #1 %
\futurelet\reserved@a\@xhline}
\makeatother

\newtheorem{definition}{Definition}

\newtheorem{problem}{Problem}
\newtheorem{optprob}{Optimization Problem}

\usetikzlibrary{arrows,arrows.meta,shapes,positioning,calc,decorations.pathreplacing,calligraphy}
\tikzstyle{smallCircle} = [draw, circle, fill=black,inner sep=0pt,outer sep=0.7pt,minimum size=2pt]
\tikzstyle{block} = [draw, rectangle,minimum height=3em, minimum width=3em]
\tikzstyle{sum} = [draw, circle]
\tikzstyle{box}=[rectangle, fill=gray!20, draw, minimum width=1.2cm, minimum height=0.5cm, align=center]
\tikzset{vertex/.style = {shape=circle,draw,minimum size=1.5em,inner sep=0}}
\tikzset{vertexFill/.style = {shape=circle,fill=gray!40!white,draw,minimum size=1.5em,inner sep=0}}
\tikzset{vertexFillMarked/.style = {shape=circle,fill=orange!50!white,draw,minimum size=1.5em,inner sep=0}}
\tikzset{edge/.style = {->,> = latex'}}
\tikzset{edgeMarked/.style = {->,> = latex',thick,darkOrange}}
\tikzset{lightning bolt to/.style={to path={
			let \p1=(\tikztostart), \p2=(\tikztotarget), \n1={veclen(\y2-\y1,\x2-\x1)} in
			(\p1) -- ($($(\p1)!0.6!(\p2)$)!\n1*.1!-90:(\p2)$) -- ($(\p1)!0.55!(\p2)$) --
			(\p2) -- ($($(\p1)!0.4!(\p2)$)!\n1*.1!90:(\p2)$) -- ($(\p1)!0.45!(\p2)$) -- 
			cycle (\p2)%
}}}

\definecolor{TolBlue}   {HTML}{0077BB}
\definecolor{TolCyan}   {HTML}{33BBEE}
\definecolor{TolTeal}   {HTML}{009988}
\definecolor{TolOrange} {HTML}{EE7733}
\definecolor{TolRed}    {HTML}{CC3311}
\definecolor{TolMagenta}{HTML}{EE3377}
\definecolor{TolGrey}   {HTML}{BBBBBB}

\colorlet{blue}   {TolBlue}
\colorlet{green}  {TolTeal}
\colorlet{red}    {TolRed}
\colorlet{orange} {TolOrange}
\colorlet{grey}   {TolGrey}
\colorlet{magenta}{TolMagenta}
\colorlet{cyan}   {TolCyan}

\newcommand{\pmiss}{{P_\mathrm{miss}}}

\newcommand{\ctlu}{{u}} %
\newcommand{\pang}{{\theta}} %
\newcommand{\pvel}{{\dot{\pang}}} %
\newcommand{\bvel}{{\dot{\rho}}} %

\newcommand{\percupright}{\mathcal{J}_{\pang}}
\newcommand{\errsqupvel}{\mathcal{J}_{\pvel}}
\newcommand{\errsqubvel}{\mathcal{J}_{\bvel}}

\newcommand{\diagdots}[3][-25]{%
	\rotatebox{#1}{\makebox[0pt]{\makebox[#2]{\xleaders\hbox{$\cdot$\hskip#3}\hfill\kern0pt}}}%
}
\DeclarePairedDelimiter\abs{\lvert}{\rvert}

\newcommand{\anymiss}[2]{\ensuremath{\texttt{AnyMiss}\,(#1,#2)}}

\title{A Controller Synthesis Framework\\for Weakly-Hard Control Systems}
\author{Marc Seidel$^1$, Martina Maggio$^2$, Frank Allg\"{o}wer$^1$\\[1mm]
{\small $^1\,$University of Stuttgart, Institute for Systems Theory and Automatic Control, Germany}\\[-0.5mm]
{\small $^2\,$Saarland University, Department of Computer Science, Germany \& Lund University, Department of Automatic Control, Sweden}
}

\usepackage{textcomp,hyperref}
\newcommand\copyrighttext{%
	\footnotesize \textcopyright 2026 IEEE. Personal use of this material is permitted.
	Permission from IEEE must be obtained for all other uses, in any current or future 
	media, including reprinting/republishing this material for advertising or promotional 
	purposes, creating new collective works, for resale or redistribution to servers or 
	lists, or reuse of any copyrighted component of this work in other works. 
}
\newcommand\copyrightnotice{%
	\begin{tikzpicture}[remember picture,overlay]
		\node[anchor=south,yshift=10pt] at (current page.south) {\fbox{\parbox{\dimexpr\textwidth-\fboxsep-\fboxrule\relax}{\copyrighttext}}};
	\end{tikzpicture}%
}

\begin{document}

\maketitle
\copyrightnotice

\begin{abstract}
	Deadline misses are more common in real-world systems than one may expect.
	The weakly-hard task model has become a standard abstraction to describe and analyze how often these misses occur, and has been especially used in control applications.
	Most existing control approaches check whether a controller manages to stabilize the system it controls when its implementation occasionally misses deadlines. However, they usually do not incorporate deadline-overrun knowledge during the controller synthesis process.
	In this paper, we present a framework that explicitly integrates weakly-hard constraints into the control design.
	Our method supports various overrun handling strategies and guarantees stability and performance under weakly-hard constraints.
	We validate the synthesized controllers on a Furuta pendulum, a representative control benchmark.
	The results show that constraint-aware controllers significantly outperform traditional designs, demonstrating the benefits of proactive and informed synthesis for overrun-aware real-time control.
\end{abstract}

\section{Introduction}

In real-time control systems, deadline misses are an unavoidable reality, as confirmed by surveys conducted among industrial practitioners~\cite{Akesson2020,Akesson2022}.
In fact, occasional timing violations are not only common but often tolerated in real systems~\cite{Cervin2005}.
Despite the best efforts in scheduling and resource provisioning, transient overloads, computational delays, or communication latencies can lead to tasks missing their deadlines~\cite{Maggio2020}.

To capture the patterns of deadline misses that may occur in real systems, both the industrial and academic communities have converged on the weakly-hard (WH) task model~\cite{Bernat2001a, Hamdaoui1995}.
Rather than requiring strict adherence to every deadline, WH models allow a bounded number of misses over a finite sequence of activations, typically expressed as the \anymiss{r}{s} constraint, meaning that at most $r$ deadline misses may occur in any window of $s$ consecutive activations.
This model offers a realistic abstraction of system behavior via finite state machines~\cite{Vreman2022c}, while still enabling formal reasoning about scheduling~\cite{Xu2023, Yeolekar2025}, stability, and control performance~\cite{Lang2025, Seidel2024b, Horssen2016}.

Research on weakly-hard real-time control has largely focused on analysis or synthesis for stabilization.
This means checking whether a controller designed for nominal timing conditions remains stable under a given WH execution pattern~\cite{Maggio2020, Vreman2022b} or designing a controller that can stabilize a system under WH packet dropouts~\cite{Horssen2016, Blind2015, Linsenmayer2017, Linsenmayer2021a, Ghosh2018}.
Techniques such as Lyapunov functions and the joint spectral radius have proven effective in establishing guarantees of stability and bounded performance degradation~\cite{Horssen2016, Vreman2021, Pazzaglia2018}.
However, the synthesis approaches are usually limited to packet dropouts and the analysis approaches typically treat deadline misses as an afterthought: the controller is designed without considering them, and robustness to deadline misses is verified only afterward.

This paper takes a different perspective.
We ask a fundamental question: What if we know in advance that certain patterns of deadline misses will occur?
Can we use this knowledge proactively for controller synthesis to achieve better performance and robustness?
Our work addresses this open problem by introducing a controller synthesis framework that explicitly incorporates WH constraints during the controller design phase.

To ground our approach in a meaningful real-world scenario, we implement and test our controllers on a Furuta pendulum, a benchmark system that exhibits many of the challenges faced in controller design and has been used as a representative control example in recent work~\cite{Braun2025, Nauta2025}. %
These include intrinsic physical instability, complex nonlinear dynamics that must be linearized for control, and multiple sensor inputs that must be fused in real time.

Our experimental results show that controllers designed with awareness of WH constraints can significantly outperform traditional designs.
By accounting for the structure of deadline miss-patterns, we are able to maintain system stability and performance even under aggressive timing faults. This highlights the value of shifting from reactive analysis to proactive (constraint-aware) synthesis in the design of robust real-time control systems.

The contribution of this paper is as follows:
\begin{itemize}
	\item[--] A controller synthesis framework for control tasks under WH constraints that explicitly supports multiple deadline handling strategies. Our work is inspired by prior work that focuses on packet dropouts in networked control systems~\cite{Linsenmayer2017, Linsenmayer2021a, Seidel2024b}, but our framework incorporates the semantics of the deadline handling strategies directly into the controller synthesis phase. This allows the resulting controller to adapt to deadline overruns.
	\item[--] A synthesis method based on switched system representations and linear matrix inequalities. Our method accounts for the structure of WH constraints and generates both non-switching and switching controllers that guarantee stability and performance, both robust with respect to a given pattern of deadline misses. The switching controller can adjust its behavior based on recent execution history, improving robustness to larger numbers of deadline misses.
	\item[--] A comprehensive experimental evaluation on a Furuta pendulum, a classic control benchmark. We demonstrate that constraint-aware controllers designed using our framework significantly outperform traditional designs in terms of stability and control performance. These results highlight the practical impact of incorporating deadline miss patterns and overrun semantics into the control design process.
\end{itemize}

The remainder of this paper is organized as follows.
Section~\ref{sec:related} introduces related work.
Section~\ref{sec:background} provides background on control theory and WH tasks that are needed in the following sections.
Section~\ref{sec:controldesign} presents the control problems and solutions and Section~\ref{sec:evaluation} shows our experimental evaluation.
We conclude with Section~\ref{sec:concl}.

\section{Related Work}
\label{sec:related}

Traditional real-time analysis often assumes that every deadline miss or packet loss is catastrophic.
This assumption is overly rigid for many control systems, where occasional faults can be tolerated provided they are sufficiently rare or bounded~\cite{Bernat2001a, Davis2019, Bruggen2021, Bozhko2021}.
To address this, researchers have introduced more flexible failure models:

\begin{itemize}
\item[--] \emph{Probabilistic} approaches describe failures as stochastic processes (for example, using Markov models to represent mode switches between ``healthy'' and ``faulty'' operation) providing guarantees in terms of risk rather than absolute bounds~\cite{Davis2019b, Davis2019, Markovic2024, Costa2005, Antunes2021}.

\item[--] \emph{Deterministic} approaches, such as the WH task model~\cite{Bernat2001a}, capture bounded sequences of failures by expressing them as constraints on periodic operations.
This model has been successfully applied to control system verification~\cite{Hammadeh2014, Blind2015, Xu2015, Ahrendts2017, Hammadeh2017, Linsenmayer2017, Ghosh2018, Pazzaglia2018, Gujarati2019, Hammadeh2020, Wu2020, Hertneck2021, Linsenmayer2021a, Hobbs2022, Xu2023} and even adopted in industrial settings~\cite{Maggio2020, Gallant2024, Gallant2025}.
\end{itemize}

From an analysis and controller synthesis perspective, the closest contributions are related to deterministic models and fault-tolerant systems.
They have been investigated in theory and motivated using networked control settings.
However, the theory of this setting is limited in its applicability to real-time control systems to specific implementations (namely tasks that are using the \emph{Kill} overrun strategy, described in Section~\ref{sec:background}), since a packet dropout due to lossy communication equals (in view of the underlying dynamics) a job that is terminated when the deadline is missed.
Early work focused on models that bound the maximum number of consecutive dropouts, developing both analysis~\cite{Schendel2010} and controller synthesis techniques~\cite{Xiong2007,GarciaRivera2007}.
There are also methods for linear systems with data loss under an \anymiss{r}{s} for analysis \cite{Jia2005,Horssen2016} and synthesis \cite{Horssen2016}.
Later, WH constraints have been used to model packet dropouts in networked control settings, generalizing above mentioned dropout models.
This includes stability analysis~\cite{Blind2015,Linsenmayer2017,Vreman2022b}, also for nonlinear systems~\cite{Hertneck2020, Hertneck2021}, and stability and performance-based controller synthesis~\cite{Seidel2024b, Lang2025, Gaid2008, Linsenmayer2017, Linsenmayer2021a}.

From this we conclude that there are several contributions that offer theoretical guarantees (and synthesis methods), but they all consider simplifying assumptions on the execution in a real-time control setting.
In particular, most existing approaches to control design for WH systems implicitly assume a kill-task semantics: at a deadline miss, the controller is reset and any partial computation is discarded (e.g., the approaches in~\cite{Blind2015, Linsenmayer2017, Vreman2022b, Seidel2024b, Lang2025} and related works).
This assumption is convenient because the analysis and synthesis methods developed for packet dropouts can then be directly applied to the deadline miss scenario, but it is not representative of typical controller implementations.
In fact, common operating systems do not support ``kill on overrun'' out of the box.
If such semantics are desired, the control engineer must implement them explicitly, e.g., by inserting deadline checks during execution and triggering a self-reset operation when the computation is unsuccessful.
While feasible, this is not common practice, and adds non-negligible computational overhead, which is undesirable precisely in settings where deadlines are already tight and misses may occur.

In many existing systems, it is much more likely that the controller continues executing and performs a time check only at the end of its computation, using that check to decide when to start the next iteration at the next natural activation point.
This execution semantic corresponds to the \emph{Skip} strategy (described in Section~\ref{sec:background}) rather than \emph{Kill}.
The approaches mentioned above for controller synthesis are therefore not explicitly applicable to cases in which tasks cannot be aborted (or are not aborted), and the controller state is not rolled back at a miss.
While some related work considers skipping behavior~\cite{Gallant2024, Gallant2025}, these approaches typically provide only probabilistic guarantees, and may entail substantial online computational overhead, making it difficult to deploy in control tasks with small periods.
The challenge of adapting these theoretic results to methods with assumptions that are closer to the controller implementation is so far widely unsolved, and this paper aims at making a contribution in that direction.

\section{Background}
\label{sec:background}

We consider the discrete-time linear time-invariant plant
\begin{equation}
	\begin{aligned}
		x(t+1) &= Ax(t) + Bu_\mathrm{a}(t) + B^w w(t)\\
		z(t) &= C x(t) + Du_\mathrm{a}(t) + D^w w(t)
	\end{aligned} \label{eq:system}
\end{equation}
with time $t \in \mathbb{N}_0$, initial state $x_0 \coloneq x(0) \in \mathbb{R}^{n}$, actuator input $u_\mathrm{a}(t) \in \mathbb{R}^{m}$, performance input $w(t) \in \mathbb{R}^{q}$, performance output $z(t) \in \mathbb{R}^{p}$, and system matrices of appropriate dimensions with real-number entries.
The objective of this work is to design a control law and implement it in a control task $\mathcal{T}$.
Said task executes according to the \emph{Logical Execution Time} paradigm \cite{Kirsch2011}, and hence the corresponding control law should consider that the controller implementation introduces a 1-step delay~\cite{Cervin1999}.
To that end, we introduce the delayed plant model
\begin{equation}
	\begin{aligned}
			\begin{bmatrix}
				x(t+1) \\ u(t+1)
			\end{bmatrix} &= \begin{bmatrix}
				A & B \\
				0 & 0
			\end{bmatrix} \! \begin{bmatrix}
				x(t) \\ u(t)
			\end{bmatrix} + \begin{bmatrix}
				0 \\ I
			\end{bmatrix}u_\mathrm{a}(t) + \begin{bmatrix}
				B^w \\ 0
			\end{bmatrix} \! w(t)\\
			z(t) &= \begin{bmatrix}
				C & D
			\end{bmatrix} \! \begin{bmatrix}
			x(t) \\ u(t)
			\end{bmatrix} + 0 \cdot u_\mathrm{a}(t) + D^w w(t)
	\end{aligned} \label{eq:system-delayed}
\end{equation}
with the augmented state $u(t) = u_\mathrm{a}(t-1) \in \mathbb{R}^{m}$ and some initial value $u(0)$.
Here, $I$ and $0$ denote the identity and zero matrix of suitable dimension, respectively.
The control task computes stateless state-feedback controllers of the form
\begin{equation}
	u_\mathrm{c}(t) = K \begin{bmatrix}
	    x(t)^\top & u(t)^\top
	\end{bmatrix}^\top, \label{eq:controller}
\end{equation}
with the computed control input $u_\mathrm{c}(t) \in \mathbb{R}^m$ and the controller gain matrix $K \in \mathbb{R}^{m \times n}$.
Note that the inherent one-step delay is already contained in the delayed plant model~\eqref{eq:system-delayed}.
Further, note the difference between the applied input $u_\mathrm{a}(t)$ and the computed input $u_\mathrm{c}(t)$.
In the next subsection we discuss how $u_\mathrm{a}(t)$ and $u_\mathrm{c}(t)$ relate to one another.

\subsection{Modeling the Effects of Deadline Misses}
\label{sec:background-lossModel}
Under nominal conditions, all activated jobs of the control task $\mathcal{T}$ finish before their deadlines.
Then, the applied control signal $u_\mathrm{a}(t)$ equals the computed control signal $u_\mathrm{c}(t)$, i.e., $u_\mathrm{a}(t) = u_\mathrm{c}(t)$.
Hence, the controller~\eqref{eq:controller} may directly influence the (delayed) plant at any time $t$.\footnote{Note that this corresponds to the logical execution time paradigm application, due to the delay being applied at the plant level.}
However, some control tasks' jobs may miss their deadline.
Thus, $u_\mathrm{a}(t)$ and $u_\mathrm{c}(t)$ are not always equal.
To describe deadline misses formally, we introduce the binary \emph{hit/miss sequence} $\mu \coloneq (\mu(t))_{t \in \mathbb{N}_0}$, where $\mu(t) = 0$ in case of a deadline miss and $\mu(t) = 1$ for a deadline hit.
For the purpose of this hit/miss sequence, we interpret a ``recovery hit'' (i.e., the hit of a job that missed at least one previous deadline and was allowed to continue executing) as a hit.
This means that $\mu(t)$ encodes whether a control job has completed its execution during the $t$-th period.
Without loss of generality, we assume $\mu(0) = 1$, i.e,.~the very first job of the control task completes before its deadline.

Two immediate questions arise whenever a deadline miss occurs: (i) how to set the next control input and (ii) what to do with the control job that exhibited the computational overrun.
We use the term \emph{actuator strategy} to refer to the answer to (i) and \emph{overrun strategy} to refer to the answer to (ii).

Two commonly used actuator strategies are \emph{Zero} and \emph{Hold}.
For \emph{Zero}, the next control input is simply set to zero, while for \emph{Hold} the previous input is held until a new control input is available.
Formally, we can write
\begin{equation}
	u_\mathrm{a}(t) = \begin{cases}
		\mu(t) u_\mathrm{c}(t) &\quad \text{\emph{Zero}}\\
		\mu(t) u_\mathrm{c}(t) + (1-\mu(t))u_\mathrm{a}(t-1) &\quad \text{\emph{Hold}}
	\end{cases}
	\label{eq:actuatorStrategies}
\end{equation}
Which of those actuator strategies is more advantageous depends heavily on the control system \cite{Schenato2009}.
Therefore, both actuator strategies are practically relevant.

The literature considers mainly three overrun strategies: \emph{Kill}, \emph{Skip}, and \emph{Queue} \cite{Pazzaglia2019, Maggio2020, Cervin2005a}.
For \emph{Kill}, the job is killed at its deadline and a new one is started, effectively omitting all computations of the killed job.
The \emph{Skip} strategy on the other hand allows the job to continue after it missed its deadline and skips the next job entirely.
This allows the job to finish.
In case the next job is not skipped but queued instead, i.e., as soon as the first job finishes, the queued one starts immediately, the \emph{Queue} strategy emerges.
Note that in principle it is possible to queue multiple jobs, but in light of control tasks and stateless controllers it is only reasonable to queue at most one job, referring to as \emph{Queue$\,$(1)}.
Queuing more than one job would inherently use old state information, which is disadvantageous for the control goals.
Thus, for stateless control jobs, typically only \emph{Queue$\,$(1)} is considered~\cite{Pazzaglia2019, Cervin2005a}.

Recently, \emph{Queue} has been empirically shown to be beneficial~\cite{Braun2025}.
However, the ability to provide formal guarantees when using \emph{Queue} is impaired by the presence of different alternatives after a job overruns:
The original job could not finish, it could finish but the queued one could not, or it finished and a fresh job also completes its execution.
This makes the outcome nondeterministic~\cite{Maggio2020}, hence harder for controller synthesis purposes.
As we want formal arguments, in this paper, we focus on \emph{Kill} and \emph{Skip} only.\footnote{The paper makes two contributions: analysis and design.
It is in principle possible to analyze the \emph{Queue$\,$(1)} strategy with our proposed framework, similarly to the analysis proposed in~\cite{Maggio2020}, but the control design proves to be significantly more challenging.
This is due to the more involved dynamics that are required to adequately model the overall system under \emph{Queue$\,$(1)}.}

\subsection{Modeling the Occurrence of Deadline Misses}
\label{sec:background-systemClass}

In many practical implementations, factors such as overload are causing deadline misses.
These factors are typically known in the system but external to the control pipeline, making their occurrence to some extent predictable.
The automotive industry has \emph{de facto} adopted the WH task model~\cite{Hammadeh2017, Ahrendts2018, Maggio2020, Hammadeh2020, Gallant2025} to describe the occurrence of deadline misses, that constrains the permissible number of deadline misses within a specific time window.
\begin{definition}[WH constraint: \texttt{AnyMiss}]
	A control task $\mathcal{T}$ satisfies the WH constraint $\lambda = \anymiss{r}{s}$, denoted by $\mathcal{T} \vdash \lambda$, if in any $s$ consecutive job activations of $\mathcal{T}$, at most $r$ of them miss their deadline.
\end{definition}
For a more detailed description and other types of WH constraints, such as \texttt{AnyHit}, \texttt{RowHit}, and \texttt{RowMiss}, we refer to the seminal paper \cite{Bernat2001a}.
The framework presented in this paper is similarly applicable to other known types of WH constraints.
With a slight abuse of notation, we also write $\mu \vdash \lambda$ whenever the hit/miss sequence $\mu$ associated with $\mathcal{T}$ satisfies the WH constraint $\lambda$.

We are now ready to formally define our considered system class, namely the class of \emph{WH control systems}, as follows.

\begin{definition}[WH control system] \label{def:WHCS}
	A WH control system is the plant~\eqref{eq:system} controlled by the control task $\mathcal{T}$ satisfying the WH constraint $\lambda$ and computing the controller~\eqref{eq:controller} and either \emph{Zero} or \emph{Hold} as actuator strategy, and \emph{Kill} or \emph{Skip} as overrun strategy.
\end{definition}

The synthesis of controllers for WH control systems is the main topic of the paper at hand.
One of our control goals, apart from stability, is to synthesize controllers with guaranteed performance.
The considered performance metric is introduced in the subsequent subsection.

\subsection{Control System Performance}
\label{sec:background-performance}
When evaluating a system's behavior, merely considering stability is typically not sufficient.
Rather, performance metrics are used to quantify how well a system behaves.
For WH control systems, a common metric is found in quadratic stage costs or LQR-like measures, see, e.g.,~\cite{Horssen2016,Vreman2021}.

In this paper, we consider a different performance measure, commonly known as \emph{$\ell_2$-performance} \cite{Khalil2002}.
For that, the performance input $w$ and performance output $z$ from Equations~\eqref{eq:system} and~\eqref{eq:system-delayed} are used.
Thereby, $w \to z$ is called the \emph{performance channel}.
The signal $w$ represents factors and influences on the system that are typically undesired, but cannot be avoided, such as disturbances.
On the other hand, $z$ contains quantities that are of interest for the control objective, for example weighted costs in $x$, measuring deviation from a desired operating point.
In its essence, $\ell_2$-performance measures the influence of $w$ on $z$, considered in a quadratic fashion.
When designing controllers that improve the $\ell_2$-performance, one aims at reducing the influence of the unwanted factors on the relevant quantities.
Thus, $\ell_2$-performance quantifies how large the output $z$ will be if we know how large the input $w$ is.
We formally define $\ell_2$-performance for WH control systems as follows.

\begin{definition}[$\ell_2$-performance for WH control systems] \label{def:l2}
	Denote the set of all square-summable signals $\ell_2 = \{w \, \mid \, \sum_{t=0}^{\infty} w(t)^\top w(t) < \infty \}$.
	The WH control system is said to have $\ell_2$-performance with gain $\gamma < \infty$ if it is asymptotically stable, and for $x_0 = 0$ it holds that
	\begin{equation} \label{eq:defPerformance}
		\sum_{t=0}^{\infty} z(t)^\top z(t) < \gamma^2 \sum_{t=0}^{\infty} w(t)^\top w(t) \quad \parbox{7em}{$\forall w \in \ell_2, w \neq 0$,\\ $\forall \mu \vdash \lambda$.}
	\end{equation}
\end{definition}

Note that this includes asymptotic stability, i.e.,~whenever a WH control system has a guaranteed $\ell_2$-performance gain $\gamma < \infty$ it is also guaranteed to be asymptotically stable.

In many engineering applications, physically relevant quantities naturally appear as energy-like terms.
The summation over all squared $w(t)$, resp.~$z(t)$, in \eqref{eq:defPerformance} is often able to represent such terms, making the considered performance measure applicable to a large class of practically relevant control systems.
Consequently, the set $\ell_2$ can be interpreted as all signals $w$ with finite energy.
Since Definition~\ref{def:l2} considers one $\gamma$ for all possible $w$, c.f.~\eqref{eq:defPerformance}, $\ell_2$-performance can be practically interpreted as the \emph{worst-case energy amplification} through the performance channel.
The smallest $\gamma$ such that \eqref{eq:defPerformance} is still satisfied is called the \emph{$\ell_2$-gain} of the system.
Any $\gamma$ satisfying \eqref{eq:defPerformance} thus constitutes an upper bound on the influence (or energy amplification) from $w$ to $z$.
That means, loosely speaking, $\gamma$ quantifies by how much $w$ is amplified (or attenuated) by the system.
As a consequence, a small $\gamma$ indicates a good performance, since undesired factors have small influence on relevant quantities $z$, while a large $\gamma$ means that $w$ can have a larger influence on $z$.

Note that the signals $w$ and $z$, as well as the related matrices $B^w$, $C$, $D$, and $D^w$, can in practice often be chosen or tuned in a suitable manner, as they are typically artificial.
For example, by altering $C$ and $D$ one can tune how relevant a state or an input is for the overall behavior of the control system.
Or it might be unknown how exactly a disturbance affects the system, but from an engineering viewpoint it might be clear that it affects one state more than another.
In this case, $B^w$ can be chosen suitably to reflect this influence.
As a consequence, the considered performance measure is more flexible in practice than one might think.
We provide reasoning for choosing the performance channel based on the desired behavior of a real-world example in our experimental evaluation in Section~\ref{sec:evaluation}.

\section{Controller Synthesis}
\label{sec:controldesign}
This section introduces our control design framework.
Its main purpose is the design of controllers with theoretical performance guarantees for WH control systems.
However, our framework also allows us to analyze WH control systems under existing controllers for their stability and/or performance.
In summary, we aim to provide solutions to the following three different analysis and synthesis problems for $\ell_2$-performance of WH control systems.

\begin{problem} \label{prob:analysis}
	Analyze a given WH control system with a given controller for its guaranteed $\ell_2$-performance.
\end{problem}

\begin{problem} \label{prob:synth-nonSwitched}
	Given a WH control system, find a non-switching controller that achieves the best guaranteed $\ell_2$-performance.
\end{problem}

\begin{problem} \label{prob:synth-switched}
	Given a WH control system, find a switching controller that achieves the best guaranteed $\ell_2$-performance.
\end{problem}

Problem~\ref{prob:analysis} constitutes an analysis problem.
On the other hand, Problems~\ref{prob:synth-nonSwitched} and~\ref{prob:synth-switched} deal with the synthesis of WH control systems with guaranteed $\ell_2$-performance.
The subtle yet crucial difference between Problem~\ref{prob:synth-nonSwitched} and~\ref{prob:synth-switched} is whether the controller matrix $K$ may be switching (i.e., be different for different controller jobs) or non-switching.
We explain the functionality of a switching controller, the difference between switching and non-switching controllers, and their influence on the design procedure in Subsection~\ref{sec:controldesign-LMIproblem}.

\subsection{The framework}
\label{sec:controldesign-framework}
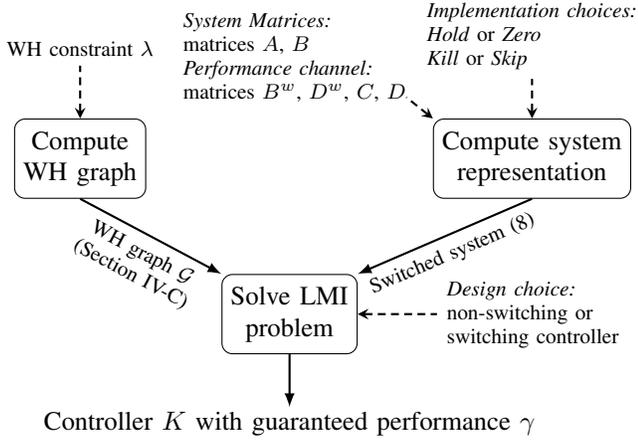
\begin{figure}[t]
	\centering
	\begin{tikzpicture}[>=stealth]
		\node[block,align=center,rounded corners] (lmi) {Solve LMI\\problem};
		\node[block,above left=of lmi,align=center,rounded corners] (graph) {Compute\\WH graph};
		\node[block,above right=of lmi,align=center,rounded corners] (rep) {Compute system\\representation};

		\draw[-latex,thick] (rep.south) -- node[midway, below, sloped] {\footnotesize Switched system~\eqref{eq:switchedSystem}} (lmi.north east);
		\draw[-latex,thick] (graph.south) -- node[midway, below, sloped, align=center, font=\footnotesize] {WH graph $\mathcal{G}$\\(Section~\ref{sec:controldesign-WHgraph})} (lmi.north west);
		\draw[-latex,thick] (lmi.south) -- node[anchor=north,pos=0.9,align=center] {Controller $K$ with guaranteed performance $\gamma$} +(0,-2em);
		
		\draw[<-,thick,densely dashed] (lmi.east) -- node[anchor=west,pos=1.0,align=left,font=\footnotesize] {\emph{Design choice:}\\non-switching or\\switching controller} +(3em,0);
		\draw[<-,thick,densely dashed] (graph.north) -- node[anchor=south,pos=1.0,align=left] {\footnotesize WH constraint $\lambda$} +(0,2em);
		\draw[<-,thick,densely dashed] (rep.north west) -- node[anchor=south east, pos=0.35, align=left, font=\footnotesize, xshift=-0.3em] {\emph{System Matrices:}\\matrices $A$, $B$\\\emph{Performance channel:}\\matrices $B^w$, $D^w$, $C$, $D$} +(-1em,0.8em);
		\draw[<-,thick,densely dashed] (rep.north) -- node[anchor=south,pos=1.0,align=left,font=\footnotesize] {\emph{Implementation choices:}\\\emph{Hold} or \emph{Zero}\\\emph{Kill} or \emph{Skip}} +(0em,1.5em);
	\end{tikzpicture}
	\caption{Workflow of the proposed controller design framework.}
	\label{fig:framework-schematics}
\end{figure}

The workflow of the proposed framework\footnote{Functional \textsc{MATLAB}-code for designing controllers using this framework is available at \texttt{\url{https://gitlab.cs.uni-saarland.de/rtsc/gss}}.} for synthesizing controllers involves three different components:
\begin{enumerate}
	\item \label{framework-step-sysrep} Computing a suitable system representation,
	\item \label{framework-step-graph} Computing the WH graph,
	\item \label{framework-step-lmi} Solving a linear matrix inequality optimization problem.
\end{enumerate}
An overview of the different components, their interplay, and how the different aspects of the WH control system and design choices of the control task influence the workflow is depicted in Figure~\ref{fig:framework-schematics}.

In 1), the plant model is to be put into an adequate system representation.
To compute this representation, the performance channel and controller implementation have to be chosen.
The result is a so-called \emph{switched system}~\cite{Lin2009}, a system class that is widely used and investigated and which we introduce in Subsection~\ref{sec:controldesign-systemrep}.
In parallel, the representation of the WH constraint as a graph, the so-called \emph{WH graph}, can be computed for 2).
Together with controller design choices, a well-behaved convex optimization problem is solved as a last step.
The output is a deadline miss-aware controller with a guaranteed performance level.
In case the result is not satisfactory, one or more of the inputs to the procedure can be altered.
The involved components then need to be recomputed.

\subsection{System representation}
\label{sec:controldesign-systemrep}
The goal of this subsection is to formulate the WH control system as a mathematical model known as \emph{switched system}.
Switched systems are piecewise linear systems that at each time step may admit different dynamics.
Its general form in closed-loop is
\begin{equation}
	\begin{alignedat}{2}
		\tilde{x}(\tilde{t} + 1) &{}={}& \mathcal{A}^\mathrm{cl}_{\sigma(\tilde{t})} \tilde{x}(\tilde{t}) &+ \mathcal{B}^w_{\sigma(\tilde{t})} \tilde{w}(\tilde{t})\\
		\tilde{z}(\tilde{t}) &{}={}& \mathcal{C}^\mathrm{cl}_{\sigma(\tilde{t})} \tilde{x}(\tilde{t}) &+ \mathcal{D}^w_{\sigma(\tilde{t})} \tilde{w}(\tilde{t}),
	\end{alignedat} \label{eq:switchedSystem-general}
\end{equation}
where the switched system matrices denoted by calligraphic letters may switch in each time step $\tilde{t}$ between a finite set of matrices, i.e.,~$( \mathcal{A}^\mathrm{cl}_{\sigma(\tilde{t})}, \mathcal{B}^w_{\sigma(\tilde{t})}, \mathcal{C}^\mathrm{cl}_{\sigma(\tilde{t})}, \mathcal{D}^w_{\sigma(\tilde{t})} ) \in \{ ( \mathcal{A}^\mathrm{cl}_l, \mathcal{B}^w_l, \mathcal{C}^\mathrm{cl}_l, \mathcal{D}^w_l ) \, \mid \, l = 1, \dots, L < \infty\}$ for all $\tilde{t}$.
The variable $l$ is called \emph{mode} of the switched system and denotes which set of system matrices is currently active, and $\sigma(\tilde{t}) \in \{ 1, \dots, L \}$ is called the \emph{switching signal}.

To reformulate the WH control system given actuator and overrun strategies as a switched system of the form~\eqref{eq:switchedSystem-general}, it is helpful to consider only time instants at which state measurements actually influence the plant through the controller via~\eqref{eq:controller}.
Due to deadline misses, at some time instants the controller does not influence the plant and the system essentially runs open-loop.
In the context of the \emph{Kill} strategy, considering only the influential time points was previously called \emph{lifting}~\cite{Seidel2024b} or \emph{alternative discretization}~\cite{Linsenmayer2017}.

First, consider the \emph{Kill} strategy.
Since killing a job essentially deletes the state information that the job was started with, only state information at time instants where a deadline hit occurs are relevant for control input actuation.
Therefore, we only consider time points where $\mu(t) = 1$.
Formally, we define $\tau_{\textit{Kill}} = \{ t \, \mid \, \mu(t) = 1 \}$ as the set of relevant time instants.
Denote with $\tau_{\textit{Kill}}(\tilde{t})$ the $\tilde{t}$-th element of $\tau_{\textit{Kill}}$ in ascending order.

For the \emph{Skip} strategy, on the contrary, the job that was started after a hit (or recovery hit) continues until completion.
All following job activations are skipped and the corresponding state measurements are irrelevant for control input computation.
Thus, the state information at time instants after a hit (or recovery hit) are relevant for control input actuation.
As a consequence, we have $\tau_{\textit{Skip}} = \{ t \, \mid \, \mu(t-1) = 1 \}$ as the set of relevant time instants for the \emph{Skip} strategy.
For ease of presentation, we use $\tau$ as shorthand notation for either $\tau_{\textit{Kill}}$ or $\tau_{\textit{Skip}}$.

Next, for ease of notation, we introduce the sequence $\alpha$ that counts the number of consecutive deadline misses between hits or recoveries, similar to \cite{Linsenmayer2017,Seidel2024b}.
Since $\mu \vdash \lambda = \anymiss{r}{s}$, the elements of $\alpha$ are within the set $\Sigma \coloneq \{ 0, \dots, r \}$, where $r$ is the maximum number of consecutive deadline misses.
Further, the sequences $\mu$ and $\alpha$ provide equivalent representations of the deadline misses and  hits/recoveries, i.e., $\alpha \vdash \lambda \iff \mu \vdash \lambda$.

\begin{figure}
	\centering
	\begin{tikzpicture}
		\def\numAttempt{7}
		\def\spacingAxis{1.0}

		\draw[thick, ->, >=stealth] (-0.5*\spacingAxis,0) -- (\numAttempt*\spacingAxis-0.5*\spacingAxis,0);
		\node[right] at (\spacingAxis*\numAttempt-0.5*\spacingAxis,0) {$t$};

		\node[above left] at (-0.3*\spacingAxis,0.05) {$\mu=$};

		\foreach \k / \label in {0/1, 1/0, 2/0, 3/1, 4/1, 5/0, 6/1} {
			\draw[thick] (\k*\spacingAxis,0.1) -- (\k*\spacingAxis,-0.1);

			\ifnum \label=0
				\node[above] at (\k*\spacingAxis,0.6) {M};
			\else
				\node[above] at (\k*\spacingAxis,0.6) {H/R};
			\fi
			\node[above] at (\k*\spacingAxis,0.1) (attempt\k) {\textsf{\label}};
		}

		\node[below left] at (-0.3*\spacingAxis,-0.2) {$\tau_\textit{Kill}=$};
		\foreach \k in {0, 3, 4, 6} {
			\node[below] at (\k*\spacingAxis,-0.15) {$\k$};
		}
		\foreach \k / \size in {0/3, 3/1, 4/2, 6/0} {
			\ifnum \size=0
				\draw[] (\k*\spacingAxis-0.15,-0.3) -- (\k*\spacingAxis-0.15,-0.2) -- (\k*\spacingAxis+0.2,-0.2);
			\else
				\draw[] (\k*\spacingAxis-0.15,-0.3) -- (\k*\spacingAxis-0.15,-0.2) -- (\k*\spacingAxis+\size*\spacingAxis-0.25,-0.2) -- (\k*\spacingAxis+\size*\spacingAxis-0.25,-0.3);
			\fi
		}

		\node[below left] at (-0.3*\spacingAxis,-0.8) {$\tau_\textit{Skip}=$};
		\foreach \k in {1, 4, 5, 7} {
			\node[below] at (\k*\spacingAxis,-0.75) {$\k$};
		}
		\foreach \k / \size in {1/3, 4/1, 5/2, 7/0} {
			\ifnum \size=0
				\draw[] (\k*\spacingAxis-0.15,-0.9) -- (\k*\spacingAxis-0.15,-0.8) -- (\k*\spacingAxis+0.2,-0.8);
			\else
				\draw[] (\k*\spacingAxis-0.15,-0.9) -- (\k*\spacingAxis-0.15,-0.8) -- (\k*\spacingAxis+\size*\spacingAxis-0.25,-0.8) -- (\k*\spacingAxis+\size*\spacingAxis-0.25,-0.9);
			\fi
		}

		\node[below left] at (-0.3*\spacingAxis,-1.35) {$\alpha=$};
		\foreach \k / \label in {1.5/2, 3.5/0, 5/1} {
			\node[below] at (\k*\spacingAxis,-1.3) {$\label$};
		}
		
	\end{tikzpicture}
	\caption{Example sequences, denoting M, H, R for a miss, hit, or recovery hit, respectively.}
	\label{fig:sequences}
\end{figure}
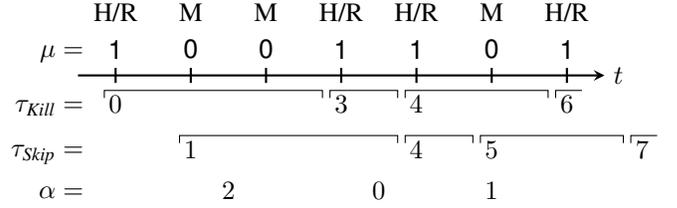

The different sequences and time instant sets $\tau$ are visualized by the following example and the corresponding Figure~\ref{fig:sequences}.
Consider the hit/miss sequence $\mu = (1 \, 0 \, 0 \, 1 \, 1 \, 0 \, 1 \dots)$ starting at time $t = 0$.
It follows that for this specific $\mu$, $\tau_{\textit{Kill}} = (0 \, 3 \, 4 \, 6 \dots)$ and $\tau_{\textit{Skip}} = (1 \, 4 \, 5 \, 7 \dots)$.
The corresponding sequence of consecutive misses is $\alpha = (2 \, 0 \, 1 \, \dots)$.

We are now ready reformulate the WH control system as a switched system of the form~\eqref{eq:switchedSystem-general}.
First, we set $\tilde{x} = [x{^\top} \,\,\, u{^\top}]{^\top} \in \mathbb{R}^{n+m}$, as we use the delayed plant model in Equation~\eqref{eq:system-delayed}.
Furthermore, since the $\tau$-sets are directly linked to time instants with hits or recovery hits, the dynamics between $\tau(\tilde{t})$ and $\tau(\tilde{t} + 1)$ depend on the corresponding value of $\alpha(\tilde{t})$, as this sequence indicates the number of misses between $\tau(\tilde{t})$ and $\tau(\tilde{t} + 1)$.
Consequently, the modes of the switched system are defined by $\sigma(\tilde{t}) = \alpha(\tilde{t}) \in \Sigma$ and correspond to the number of consecutive deadline misses between $\tau(\tilde{t})$ and $\tau(\tilde{t} + 1)$.
Thus, $\alpha$ acts as the switching sequence for \eqref{eq:switchedSystem-general}.
Since for $\ell_2$-performance, $w(t)$ and $z(t)$ for all $t$ need to be considered, c.f.\ \eqref{eq:defPerformance}, it is not sufficient to merely take $w(\tau(\tilde{t}))$ and $z(\tau(\tilde{t}))$ into account.
We therefore need to stack the performance input and output signals between those time instants, as in \cite{Seidel2024b}:
\begin{equation}
	\begin{aligned}
			\tilde{w}(\tilde{t}) = \begin{bmatrix}
				w(\tau(\tilde{t}-1)) \\ \vdots \\ w(\tau(\tilde{t})-1)
			\end{bmatrix}, \quad
			\tilde{z}(\tilde{t}) = \begin{bmatrix}
				z(\tau(\tilde{t}-1)) \\ \vdots \\ z(\tau(\tilde{t})-1)
			\end{bmatrix}.
	\end{aligned} \label{eq:lifting-performanceChannel}
\end{equation}
We then arrive at the switched system form~\eqref{eq:switchedSystem-general}, reading as
\begin{equation}
	\begin{aligned}
			\begin{bmatrix}
				x(\tilde{t}+1) \\ u(\tilde{t}+1)
			\end{bmatrix} &= (\mathcal{A}_{\alpha(\tilde{t})} + \mathcal{B}_{\alpha(\tilde{t})} K) \begin{bmatrix}
				x(\tilde{t}) \\ u(\tilde{t})
			\end{bmatrix} + \mathcal{B}^w_{\alpha(\tilde{t})} \tilde{w}(\tilde{t})\\
			\tilde{z}(\tilde{t}) &= (\mathcal{C}_{\alpha(\tilde{t})} + \mathcal{D}_{\alpha(\tilde{t})} K) \begin{bmatrix}
				x(\tilde{t}) \\ u(\tilde{t})
			\end{bmatrix} + \mathcal{D}^w_{\alpha(\tilde{t})} \tilde{w}(\tilde{t}).
	\end{aligned} \label{eq:switchedSystem}
\end{equation}
Note that due to the considerations of time instants $\tau(\tilde{t})$ only, one time step of the switched system~\eqref{eq:switchedSystem} may correspond to multiple time instants of the original plant~\eqref{eq:system}.
The matrices $\mathcal{A}_{\alpha(\tilde{t})}$, $\mathcal{B}_{\alpha(\tilde{t})}$, $\mathcal{B}^w_{\alpha(\tilde{t})}$, $\mathcal{C}_{\alpha(\tilde{t})}$, $\mathcal{D}_{\alpha(\tilde{t})}$, and $\mathcal{D}^w_{\alpha(\tilde{t})}$ depend on the chosen actuator and overrun strategies.
Through algebraic computation of the dynamics from time $\tau(\tilde{t})$ to $\tau(\tilde{t}+1)$, i.e., from one time instant that influences the system through the controller to the next one, one obtains the switched system matrices shown in Figure~\ref{fig:matrices-strategies} for~\eqref{eq:switchedSystem}.%

\begin{figure*}
\centering
\begin{tikzpicture}
\draw[densely dashed] (0,0) -- (17.5,0);
\draw[densely dashed] (0,-10) -- (17.5,-10);
\draw[densely dashed] (0,-5) -- (17.5, -5);
\draw[densely dashed] (0.5,0.25) -- (0.5,-10.25);
\draw[densely dashed] (8,0.25) -- (8,-10.25);
\draw[densely dashed] (12,0.25) -- (12,-10.25);

\node[anchor=south,rotate=90] at (0.5,-2.5) {Kill};
\node[anchor=south,rotate=90] at (0.5,-7.5) {Skip};
\node[anchor=south] at (4.25,0) {Common};
\node[anchor=south] at (10,0) {Zero};
\node[anchor=south] at (14.75,0) {Hold};

\node at (4.25,-1) {$\mathcal{A}_{\alpha} = \begin{bmatrix}
					A^{\alpha + 1} & A^{\alpha} B \\
					0 & 0
				\end{bmatrix},\,\,\,
                \mathcal{B}_{\alpha = 0} = \begin{bmatrix}
					0 \\
					I
				\end{bmatrix}$ };
\node at (4.25,-2) {$\mathcal{C}_{\alpha = 0} = \begin{bmatrix}
					C & D
				\end{bmatrix},\,\,\,
                \mathcal{D}_{\alpha = 0} = 0$};
\node at (4.25,-3.5) {$\mathcal{C}_{\alpha \geq 1} = \begin{bmatrix}
					CA^0 & D \\
					CA^1 & CA^0B \\
					\vdots & \vdots\\
					CA^{\alpha} & CA^{\alpha - 1}B
				\end{bmatrix},\,\,\,
                \mathcal{D}_{\alpha = 1} = \begin{bmatrix}
					0 \\ D
				\end{bmatrix}
                $};
\node at (10,-1) {$\mathcal{B}_{\alpha \geq 1} = \begin{bmatrix}
					A^{\alpha-1} B \\
					0
				\end{bmatrix}$};
\node at (10,-3.25) {$\mathcal{D}_{\alpha \geq 2} = \begin{bmatrix}
					0 \\ D \\ CA^0B \\ \vdots \\ CA^{\alpha - 2}B
				\end{bmatrix}$};

\node at (14.75,-1) {$\mathcal{B}_{\alpha \geq 1} = \begin{bmatrix}
					\sum\limits_{i=0}^{\alpha-1}A^i B \\
					I
				\end{bmatrix}$};
\node at (14.75,-3.25) {$\mathcal{D}_{\alpha \geq 2} = \begin{bmatrix}
					0 \\ D \\ D+CB \\ \vdots \\ D + C\sum\limits_{i=0}^{\alpha - 2}A^i B
				\end{bmatrix}$};
                
\node at (4.25,-6) {$\mathcal{B}_{\alpha} = \begin{bmatrix} 0 \\ I \end{bmatrix}, \,\,\, \mathcal{D}_{\alpha} = 0$};

\node[align=center] at (10,-6) {$\mathcal{A}_{\alpha}$ and $\mathcal{C}_{\alpha}$: same as\\Kill, Common };

\node at (14.75,-6) {$\mathcal{A}_{\alpha} = \begin{bmatrix}
					A^{\alpha + 1} & \sum\limits_{i=0}^{\alpha}A^i B \\
					0 & 0
				\end{bmatrix}$};
\node at (14.75,-7) {$\mathcal{C}_{\alpha = 0} = \begin{bmatrix}
					C & D
				\end{bmatrix}$};
\node at (14.75,-8.5) {$\mathcal{C}_{\alpha \geq 1} = \begin{bmatrix}
					CA^0 & D \\
					CA^1 & D+CB \\
					\vdots & \vdots\\
					CA^{\alpha} & D + C \sum\limits_{i=0}^{\alpha - 1}A^i B
				\end{bmatrix}$};

\node[draw, thick, rectangle, rounded corners, fill=black!10, align=left] at (6,-8.75) {
For all strategy combinations:\\[2mm]
$
\begin{aligned}
			\mathcal{B}^w_{\alpha} &= \begin{bmatrix}
				A^{\alpha} B^w & \hdots & A^0 B^w \\
				0 & \hdots & 0
			\end{bmatrix}, \quad
			\mathcal{D}^w_{\alpha = 0} = D^w \\[2mm]
			\mathcal{D}^w_{\alpha \geq 1} &= \begin{bmatrix}
				D^w & 0 & \cdots & 0 \\
				C A^0 B^w & \multicolumn{2}{c}{\smash{\raisebox{-0.8\normalbaselineskip}{\diagdots[-30]{6.70em}{.5em}}}} & \vdots \\
				\vdots & \multicolumn{1}{c}{\smash{\raisebox{0.2\normalbaselineskip}{\diagdots[-30]{4.0em}{.5em}}}} & \multicolumn{1}{c}{\smash{\raisebox{1.7\normalbaselineskip}{\diagdots[-30]{4.0em}{.5em}}}} & 0 \\
				C A^{\alpha-1} B^w & \mathclap{\cdots} & C A^0 B^w & D^w
			\end{bmatrix}
		\end{aligned} %
$
};
\end{tikzpicture}
\caption{Switched system matrices for different actuator and overrun strategy combinations, omitting the time argument of $\alpha(\tilde{t})$ for conciseness of notation.}
\label{fig:matrices-strategies}
\end{figure*}

In summary, the switched system as a result of the framework's component~\ref{framework-step-sysrep}) can be obtained by computing the matrices in Figure~\ref{fig:matrices-strategies} for the respective actuator and overrun strategies.
The system representation of the WH control system is then given by the switched system~\eqref{eq:switchedSystem}.
Recall that the switching of~\eqref{eq:switchedSystem} admits to the WH constraint.
As a next step, we aim to represent the WH constraint by a graph that generates the permissible switching signals, constituting component~\ref{framework-step-graph}) of the framework.

\subsection{Weakly-hard graph}
\label{sec:controldesign-WHgraph}
In the previous subsection, we reformulated the WH control system as a switched system with the switching signal $\alpha$.
Even though there is vast literature on switched systems with arbitrary switching, i.e.,~no restrictions on $\alpha$, the switching of \eqref{eq:switchedSystem} is in fact restricted due to the WH control system admitting to the WH constraint.
The WH constraint limits the possible hit/miss combinations and therefore imposes a switching rule on $\alpha$.
In the literature, it has been proven useful to represent the WH constraint by a graph, see, e.g., \cite{Hamdaoui1995,Blind2015,Linsenmayer2017,Vreman2022c,Seidel2024b}.
In the subsequent definition, we introduce the \emph{WH graph}, but refer to the cited literature for details.

\begin{definition}[WH graph] \label{def:WH-graph}
	Consider the directed graph $\mathcal{G} = (\mathcal{V},\mathcal{E})$, where $\mathcal{V}$ is the set of all nodes $v_i$, $i = 1, \dots, n_\mathcal{V}$, and $\mathcal{E}$ denotes the set of edges $e_k$, $k = 1, \dots, n_\mathcal{E}$, where $e_k = (i,j,l) \in \mathcal{E}$ if there exists an edge from node $v_i$ to node $v_j$ with label $l$.
	$\mathcal{G}$ is a WH graph for the WH control system if the language of $\mathcal{G}$ is equivalent to the set of all $\alpha$-sequences with $\alpha \vdash \lambda$.
\end{definition}

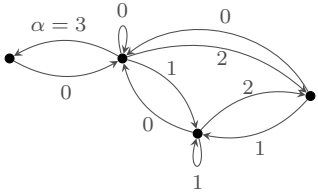
\begin{figure}
\centering
\begin{tikzpicture}[>=stealth, font=\footnotesize]
\node[draw, circle, radius=15pt, fill=black, inner sep=0pt, font=\tiny] (n1) at (0,0) {$\cdot$};
\node[draw, circle, radius=15pt, fill=black, inner sep=0pt, font=\tiny] (n2) at (-1.5,0) {$\cdot$};
\node[draw, circle, radius=15pt, fill=black, inner sep=0pt, font=\tiny] (n3) at (1,-1) {$\cdot$};
\node[draw, circle, radius=15pt, fill=black, inner sep=0pt, font=\tiny] (n4) at (2.5,-0.5) {$\cdot$};

\draw[->, black!70] (n1) to[bend right] node[above, xshift=-1mm] {$\alpha=3$} (n2);
\draw[->, black!70] (n2) to[bend right] node[below] {$0$} (n1);
\draw[->, black!70] (n1) to[bend left] node[above] {$1$} (n3);
\draw[->, black!70] (n3) to[bend left] node[below] {$0$} (n1);
\draw[->, black!70] (n1) to[bend left] node[below, yshift=0.85mm] {$2$} (n4);
\draw[->, black!70] (n4) to[bend right=50] node[above] {$0$} (n1);
\draw[->, black!70] (n3) to[bend left] node[above, yshift=-0.85mm] {$2$} (n4);
\draw[->, black!70] (n4) to[bend left] node[below] {$1$} (n3);
\draw[->, black!70] (n3) to[loop below] node[below] {$1$} (n3);
\draw[->, black!70] (n1) to[loop above] node[above] {$0$} (n1);
]
\end{tikzpicture}
\caption{Example of a WH graph $\mathcal{G}$ for the \anymiss{3}{5} constraint.}
\label{fig:example_wh_graph}
\end{figure}

All permissible hit/miss sequences satisfying the WH constraint can thus be encoded by moving along the edges of $\mathcal{G}$.
Also, all sequences $\alpha$ that correspond to a path on $\mathcal{G}$ are permissible hit/miss sequences, i.e.,~$\alpha \vdash \lambda$.
Consequently, all sequences generated by $\mathcal{G}$ constitute possible switches between the modes $\alpha(\tilde{t})$ of the switched system~\eqref{eq:switchedSystem}.
Figure~\ref{fig:example_wh_graph} shows $\mathcal{G}$ for the example of an \anymiss{3}{5} constraint.

Note that Definition~\ref{def:WH-graph} is equal to the notion of the \emph{lifted graph} from \cite{Seidel2024b} or \cite{Linsenmayer2017}, which differs slightly from earlier notions \cite{Hamdaoui1995,Vreman2022c}.
The basic principle however is the same.
The generation of such a graph is always possible and corresponding algorithms are available \cite{Linsenmayer2021a,Vreman2022c}.
This is also possible for the other types of WH constraints \cite{Bernat2001a,Vreman2022b}, which in turn means that our framework is applicable for all WH constraints.

\subsection{Solve optimization problem}
\label{sec:controldesign-LMIproblem}
Using the system representation from Subsection~\ref{sec:controldesign-systemrep} and the WH graph as explained in Subsection~\ref{sec:controldesign-WHgraph}, we can now solve Problems~\ref{prob:analysis}--\ref{prob:synth-switched}.

The solution of those problems involves solving an optimization problem containing \emph{Linear Matrix Inequalities} (LMIs).
LMIs are naturally semidefinite programs and thus convex.
They are well understood and widely used in linear control theory \cite{Boyd1994} and there are powerful tools \cite{Lofberg2004,mosek} for various programming languages for solving problems involving LMIs.
As a consequence, LMI problems can be solved reliably and efficiently.
The LMI problems appearing in the context at hand take the form 
\begin{equation}
	\begin{aligned}
		\min\limits_{\delta} \quad &\gamma \\
		\text{such that} \quad &\mathcal{M} \succ 0,
	\end{aligned} \label{eq:lmiProblem}
\end{equation}
where $\mathcal{M} = \{M_1, M_2, \dots\}$ is a set of real symmetric matrices that individually shall satisfy $M_k \succ 0$ for all $k$ and $\delta$ are the decision variables (specified later separately for each problem).
The solution of this convex optimization problem is the achieved $\ell_2$-performance gain $\gamma$ and, in case of synthesis, the optimal controller $K$ computed from the optimal values of the decision variables $\delta$.

We now specify $\mathcal{M}$ and $\delta$ for each of the Problems~\ref{prob:analysis}--\ref{prob:synth-switched} by reformulating them into the form \eqref{eq:lmiProblem}.
In all of them, the matrix set is $\mathcal{M} = \{M_k \, \mid \, e_k = (i,j,l) \in \mathcal{E}\}$, i.e., it contains $n_\mathcal{E}$ matrices $M_k$, that pose a condition on each edge $e_k$ of the WH graph $\mathcal{G} = (\mathcal{V},\mathcal{E})$.
Further, the set of decision variables contains graph-global decision variables $\delta^\mathcal{G}$ and a set of node-dependent decision variables $\delta_i$ for each $v_i \in \mathcal{V}$, i.e., $\delta = \bigcup_{v_i \in \mathcal{V}} \delta_i \cup \delta^\mathcal{G}$.
For \textit{Kill}, \cite{Seidel2024b} reports how the resulting LMI problems guarantee $\ell_2$-performance.
Here, we additionally extend this to \textit{Skip} by providing the respective switched system matrices in Figure~\ref{fig:matrices-strategies}, making it a broadly applicable framework.
Subsequently, we state the three optimization problems involving LMI problems to provide the solution for our original research Problems~\ref{prob:analysis}--\ref{prob:synth-switched} in order.

\begin{optprob}[Analysis]\label{optprob:analysis}
	The analysis Problem~\ref{prob:analysis} is solved by the LMI problem~\eqref{eq:lmiProblem} with $\delta^\mathcal{G} = \{\gamma\}$, where $\gamma \in \mathbb{R}$, and $\delta_i = \{S_i, G_i\}$, where $S_i \in \mathbb{R}^{(n+m) \times (n+m)}$ symmetric, $G_i \in \mathbb{R}^{(n+m) \times (n+m)}$, $i = 1,...,n_\mathcal{V}$, and $\mathcal{M} = \{M_k \, \mid \, e_k = (i,j,l) \in \mathcal{E}\}$ with
	\begin{equation} \label{eq:lmiAnalysis} \raisetag{1\baselineskip}
		M_k = \begin{bmatrix}
			G_i + G_i^\top - S_i & \ast & \ast & \ast \\
			(\mathcal{A}_l + \mathcal{B}_l K) G_i & S_j & \ast & \ast \\
			0 & (\mathcal{B}_l^w)^\top & \gamma I & \ast \\
			(\mathcal{C}_l + \mathcal{D}_l K) G_i & 0 & \mathcal{D}_l^w & \gamma I
		\end{bmatrix},
	\end{equation}
	where $\ast$ denote matrix blocks than result from the symmetry constraint.
\end{optprob}

The theoretical guarantee can be obtained by similar arguments as in \cite[Theorem~2]{Seidel2024b} using an equivalent congruence transformation of the involved LMIs and leading to a node-dependent Lyapunov function characterized by the $S_i$'s.

For controller synthesis, we obtain the following results.

\begin{optprob}[Non-switching controller synthesis] \label{optprob:synthesis-nonSwitched}
	The non-switching controller synthesis Problem~\ref{prob:synth-nonSwitched} is solved by the LMI problem~\eqref{eq:lmiProblem} with $\delta^\mathcal{G} = \{G, R, \gamma\}$, where $\gamma \in \mathbb{R}$, $G \in \mathbb{R}^{(n+m) \times (n+m)}$, $R \in \mathbb{R}^{m \times (n+m)}$, and $\delta_i = \{S_i\}$, where $S_i \in \mathbb{R}^{(n+m) \times (n+m)}$ symmetric, $i = 1,...,n_\mathcal{V}$, and $\mathcal{M} = \{M_k \, \mid \, e_k = (i,j,l) \in \mathcal{E}\}$ with
	\begin{equation}
		M_k = \begin{bmatrix}
			G + G^\top - S_i & \ast & \ast & \ast \\
			\mathcal{A}_l G + \mathcal{B}_l R & S_j & \ast & \ast \\
			0 & (\mathcal{B}_l^w)^\top & \gamma I & \ast \\
			\mathcal{C}_l G + \mathcal{D}_l R & 0 & \mathcal{D}_l^w  & \gamma I
		\end{bmatrix}. \label{eq:lmiSynthesis-nonSwitched} \raisetag{1\baselineskip}
	\end{equation}
	The non-switching controller is computed from the solution of LMI problem~\eqref{eq:lmiProblem} by $K = R G^{-1}$.
\end{optprob}

Note that Optimization Problem~\ref{optprob:synthesis-nonSwitched} gives us the best possible controller $K$ that guarantees the performance level $\gamma$ under any permissible hit/miss sequence.
However, the solution $K$ is then independent of the hit/miss sequence.

It is advantageous to let the controller gain matrix adapt to the current operating conditions, i.e., the currently experienced hit/miss pattern.
This is achieved by synthesizing one separate gain matrix $K_i$ per node $v_i$ of the WH graph.
Then, the applied controller is $u_\mathrm{c}(t) = K_i x(t)$ and depends on the current node of the WH graph, thus implicitly depending on the past hit/miss pattern.
This typically increases the feasibility of the approach and improves the achieved performance guarantee.
The subsequent optimization problem states the corresponding LMIs for switching controller synthesis.

\begin{optprob}[Switching controller synthesis] \label{optprob:synthesis-switched}
	The switching controller synthesis Problem~\ref{prob:synth-switched} is solved by the LMI problem~\eqref{eq:lmiProblem} with $\delta^\mathcal{G} = \{\gamma\}$, where $\gamma \in \mathbb{R}$, and $\delta_i = \{S_i, G_i, R_i\}$, where $S_i \in \mathbb{R}^{(n+m) \times (n+m)}$ symmetric, $G_i \in \mathbb{R}^{(n+m) \times (n+m)}$, $R_i \in \mathbb{R}^{m \times (n+m)}$, $i = 1,...,n_\mathcal{V}$, and $\mathcal{M} = \{M_k \, \mid \, e_k = (i,j,l) \in \mathcal{E}\}$ with
	\begin{equation}
		M_k = \begin{bmatrix}
			G_i + G_i^\top - S_i & \ast & \ast & \ast \\
			\mathcal{A}_l G_i + \mathcal{B}_l R_i & S_j & \ast & \ast \\
			0 & (\mathcal{B}_l^w)^\top & \gamma I & \ast \\
			\mathcal{C}_l G_i + \mathcal{D}_l R_i & 0 & \mathcal{D}_l^w & \gamma I
		\end{bmatrix}. \label{eq:lmiSynthesis-switched} \raisetag{1\baselineskip}
	\end{equation}
	The switching controller is computed from the solution of LMI problem~\eqref{eq:lmiProblem} by $K_i = R_i G_i^{-1}$ for each node $v_i \in \mathcal{V}$.
\end{optprob}

Finally, it is worth mentioning that one can adapt the LMIs to neglect performance and focus on stability only.
The result is either an analogous analysis procedure for asymptotic stability in the case of \eqref{eq:lmiAnalysis}, or the synthesis of a stabilizing controller without any performance considerations for~\eqref{eq:lmiSynthesis-nonSwitched} and~\eqref{eq:lmiSynthesis-switched}.
To do so, only consider the upper left 2-by-2 block of the matrices $M_k$ in~\eqref{eq:lmiAnalysis}, \eqref{eq:lmiSynthesis-nonSwitched}, or~\eqref{eq:lmiSynthesis-switched}.
Then, the LMIs collapse to the ones derived in the context of networked control systems under WH transmission dropout descriptions in \cite{Linsenmayer2017}.
Recall though that asymptotic stability is implicitly included in the definition of $\ell_2$-performance and therefore the resulting controllers will also asymptotically stabilize the WH control system.

\subsection{Discussion}
\label{sec:controldesign-discussion}
First, we want to stress that our method explicitly takes deadline misses into account, i.e.,~it allows us to design controllers specifically for a certain WH constraint.
Next, note that upon changes, the framework only requires to recalculate the necessary parts.
For example, when changing from \emph{Kill} to \emph{Skip} one only needs to recompute the system representation, but not the WH graph, as seen in Figure~\ref{fig:framework-schematics}.

Note that any obtained performance guarantee transfers to other WH control systems with a harder WH constraint~\cite{Bernat2001a,Vreman2022a}.
In fact, our framework is not limited to WH constraints: it can be used for any graph-based deadline miss model, i.e., for any regular language description of the admissible hit/miss pattern.

Observe that our framework is based on the concept of \emph{multiple Lyapunov functions}, that is known to typically provide sufficient conditions only.
Consequently, the guarantees provided by our controller design framework are only sufficient as well.
Hence, if the optimization problem has no solution, we cannot guarantee stability.
However, there still might exist a controller that stabilizes the system.
Taking a practical approach, one can solve the problem using a harder WH constraint and use the resulting controller, enforcing the harder constraint at the platform level.

The WH graph can grow large, especially for increasing window lengths $s$, which imply that many hit/miss combinations are possible.
This is essentially the trade-off for designing a controller that explicitly considers all admissible hit/miss combinations.
The reduction of the computational burden that comes with increasing graph size is subject to future work.
Note however, that our presented procedure is an offline design, i.e.,~none of the framework's components are required to be computed during runtime.
This enables the controller design for larger WH graphs, since no problem of the form~\eqref{eq:lmiProblem} needs to be posed and solved online.

Moreover, note that the presented approach is able to synthesize switching stateless state-feedback controllers.
Those are more powerful than classical static state-feedback controllers, because they may additionally depend on the last applied input $u$ according to the delayed model~\eqref{eq:system-delayed}.
The last applied input is easily known to the control task $\mathcal{T}$ and grants additional flexibility to the controller design.
Additionally, using a switching controller $K_i$ allows the controller to adapt to past hit/miss patterns and implicitly to future possible evolutions thereof.
This can be interpreted as an adaptive control task and leads to increased feasibility and performance.

The presented framework can be used to quantify by how much the guaranteed performance changes if the WH control system admits to a weaker or harder WH constraint instead.
This can be used to free computational resources if for example a certain performance guarantee can also be achieved with a different WH constraint that allows more deadline misses.
Related, one can also use the procedure of \cite{Hsieh2023} to find all WH constraints for which stability or a certain $\ell_2$-performance level $\gamma$ is guaranteed.

The computational complexity and required memory both depend on the size of the WH graph.
In the following, we derive the number of nodes $n_\mathcal{V}$ and edges $n_\mathcal{E}$ of $\mathcal{G} = (\mathcal{V},\mathcal{E})$ for an \texttt{AnyMiss} constraint.
Note that \texttt{AnyHit} and \texttt{RowMiss} constraints can be equivalently reformulated into an \texttt{AnyMiss} constraint~\cite{Vreman2022a}.
For space reasons, we also refer to~\cite{Vreman2022a} for determining the graph size of \texttt{RowHit} constraints.

Determining the graph size is a matter of combinatorics.
The nodes of the WH graph encode the admissible combinations of misses and hits/recoveries of the hit/miss pattern $\mu$ within the past window of length $s$.
We first aim to find the number of edges with a specific label $0 \leq l \leq r$, denoted $n_\mathcal{E}(l)$, corresponding to the $\alpha$ sequence.
Recall that an edge with label $l$ corresponds to a hit/miss sequence of length $l+1$ with $l$ misses and one hit/recovery, c.f., Section~\ref{sec:controldesign-systemrep}.
Thus, we want to find the number of combinations of placing the remaining $r-l$ misses within the preceding $s-(l+1)$ time steps.
Thus, we have $n_\mathcal{E}(l) = \binom{s-1-l}{r-l}$, where $\binom{\cdot}{\cdot}$ denotes the binomial coefficient.
Hence, using Pascal's identity, the total number of edges is
\begin{equation}
	n_\mathcal{E} = \sum_{l=0}^{r} n_\mathcal{E}(l) = \sum_{l=0}^{r} \binom{s-l}{r-l} - \binom{s-l-1}{r-l-1} = \binom{s}{r}.
\end{equation}
The number of nodes arises directly from the fact that every node can be left with $\alpha = 0$ and hence is equal to the number of edges with label $0$, i.e., $n_\mathcal{V} = n_\mathcal{E}(0) = \binom{s-1}{r}$.
Note that the obtained graph size is smaller than the one presented in \cite{Vreman2022c}, as the graph definitions differ and the lifting implicitly reduces the number of combinations to encode.

On the topic of computational complexity of the optimization problems, let us denote the number of scalar decision variables with $\abs{\delta}$ and the number of constraints with $\abs{\mathcal{M}}$.
In the worst case, such a problem is $\mathcal{O}(\abs{\delta}^{3} \ln \abs{\delta})$ in the decision variables and $\mathcal{O}(\abs{\mathcal{M}}^{1.5} \ln \abs{\mathcal{M}})$ in the number of constraints \cite{BenTal2001}.
However, note that the optimization problems at hand have similarities to classical Lyapunov LMIs, which have been reported to be solved significantly faster than the worst case, see \cite[Section 2.4.4]{Boyd1994}.
The number of decision variables depends both on the state dimension of the delayed plant~\eqref{eq:system-delayed} $n+m$ and on $n_\mathcal{V}$.
More precisely, for the computationally most demanding Optimization Problem~\ref{optprob:synthesis-switched},
\begin{equation}
	\begin{aligned}
		\abs{\delta} ={}& n_\mathcal{V} \Big( \underbrace{(n+m)(n+m+1)/2}_{S_i} + \underbrace{(n+m)(n+m)}_{G_i} \\
		&+ \underbrace{m(m+n)}_{R_i} \Big) + \underbrace{1}_{\gamma}.
	\end{aligned}
\end{equation}
The number of constraints is exactly the number of edges of the WH graph, i.e., $\abs{\mathcal{M}} = n_\mathcal{E}$.

The code size of the designed controller depends on whether the controller is switching or non-switching.
Specifically, the switching controller requires the storage of $(n+m)$ gains (dimension of delayed plan model \eqref{eq:system-delayed}) per component of the control signal $u$ and per node of the graph.
This totals to $n_\mathcal{V} \cdot (n+m) \cdot m$ coefficients (floating points or fixed point values).
Space is a reason why the non-switching controller may be preferable in some cases, e.g., on memory-constrained systems, even though the switching controller theoretically guarantees better performance.
This is also clearly visible in practice.

\subsection{Extensions to the framework}
\label{sec:controldesign-extensions}
With a few adaptations, the presented framework is flexible and can be used for more applications.

For example, even though the framework includes only stateless controllers with gain matrices $K$ or $K_i$, it can also be used to analyze dynamic controllers.
Dynamic controllers are often investigated in many related works, for example \cite{Vreman2021}.
As long as one can find a closed-loop switched system representation~\eqref{eq:switchedSystem}, Optimization Problem~\ref{optprob:analysis} can be used for $\ell_2$-performance analysis.
Synthesizing dynamic or output-feedback controllers however is significantly more challenging and open for future research.

In this paper, we consider $\ell_2$-performance, a performance metric very commonly used in practice.
Adapting the LMIs to be solved enables to tackle also other performance metrics, such as quadratic performance~\cite{Lang2025}.
It is further possible to include robustness against uncertainties \cite{Lang2025}.
For plants with state interactions that are constrained by an underlying cone, e.g., widely used positive systems, different performance metrics can be analyzed.
In the case of positive systems, the LMIs reduce to linear programs.
Thus, our framework can cope with a broader class of systems.
For details thereon we refer to \cite{Seidel2025}.

\section{Experimental Evaluation}
\label{sec:evaluation}

\begin{figure}
	\centering
	\includegraphics[width=0.9\columnwidth, frame]{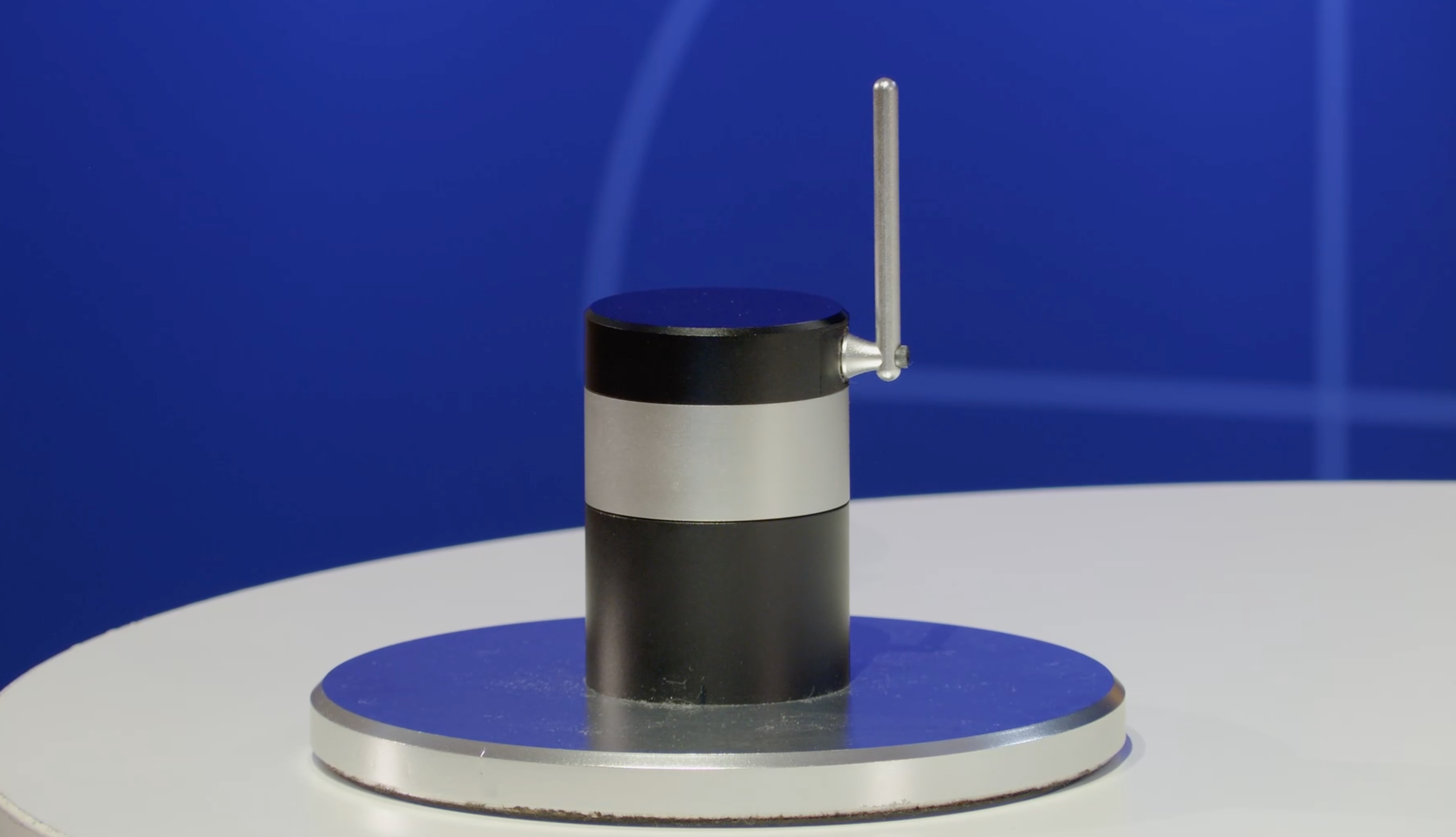}
	\caption{The Furuta pendulum used for experimental evaluation.}
	\label{fig:furuta}
\end{figure}

In this section, we present the results of the experimental evaluation conducted on a small Furuta pendulum,\footnote{We use a pendulum built according to the instructions provided in the Furuta pendulum posts of \url{https://build-its-inprogress.blogspot.com}.} depicted in Figure~\ref{fig:furuta}.
Section~\ref{sec:exp:setup} presents the experimental setup and the performance metrics used for the comparison.
Sections~\ref{sec:exp:kz} and~\ref{sec:exp:kh} present two experimental setups, comparing controllers designed respectively for the \emph{Kill and Zero} and the \emph{Kill and Hold} policy.
Sections~\ref{sec:exp:sz} and~\ref{sec:exp:sh} present the remaining two experimental setups, comparing controllers designed respectively for the \emph{Skip and Zero} and the \emph{Skip and Hold} policy.

\subsection{Experimental Setup}
\label{sec:exp:setup}
The control task executes periodically every $5\,ms$, with activations at times $a(t)$, where $t$ counts the job executions.
At time $a(t)$, the controller task applies the control signal $\ctlu(t)$ that has been calculated during the previous period, i.e., the torque applied at the base level.
It then proceeds to read data from the pendulum sensors: the pendulum angle $\pang(t)$, the pendulum angular velocity $\pvel(t)$, and the base velocity $\bvel(t)$.
The controller is also aware of its own computational state, i.e., of the sequence $(\mu_i)_{i \in \{ t-s+1, \dots, t-1 \}}$, that is the hit/miss sequences over the past window of length $s$.

The controller has three operational modes:
\begin{enumerate*}[label=(\roman*)]
	\item initialization,
	\item swing up mode,
	\item linear mode.
\end{enumerate*}
The initialization only lasts for a single iteration, in which the pendulum is forced to move by applying a small torque at the base.
The swing up mode is then designed to let the pendulum reach the proximity of the upright position (we do not inject deadline misses during the swing up phase, because this would make our experiments not comparable).
Finally, the controller operates in linear mode.
In this mode we inject deadline misses whose patterns satisfy a WH constraint.
Note that -- in principle -- each experiment tests a different randomized pattern.
If a deadline miss can occur, depending on what happened during the window, we introduce it with a certain probability.
When the probability is $1.0$, the experiments and the executions still differ due to the pendulum state in which the first deadline miss is injected.

The remaining subsections show different experimental setups that allow us to compare several controllers.
For each experimental setup and each controller, we report statistics on the execution of $100$ experiments\footnote{A video showing a subset of those experiments can be found here: \texttt{\url{https://youtu.be/dtjpJVKU5K4}}}.
For each experiment, we generate a sequence of deadline misses $(\mu_i)_{i \in \{1,\dots, 25000\}}$ that satisfies a WH constraint \anymiss{r}{s}.
For the generation, we use the value $\pmiss$ to indicate the probability of missing a deadline when possible.
This sequence corresponds to roughly $2\,$minutes of execution of the controller code (specifically $125\,$seconds; the first $5\,$seconds are considered necessary for initialization and swing up to complete, and are discounted from the performance calculation).

We compare our controllers with two baseline alternatives.
The first one is the \texttt{original} controller presented in~\cite{Josephrexon2022}, and the second one is a linear quadratic regulator \texttt{lqr} presented in~\cite{Nauta2025}.
These are defined by
\begin{align}
	\text{\texttt{original}}\!: u_\mathrm{c} &= 0.3750 \, {\pang} + 0.0250 \, \pvel + 0.0125 \, \bvel \label{eq:original} \\
	\text{\texttt{lqr}}\!: u_\mathrm{c} &= 0.4280 \, {\pang} + 0.0307 \, \pvel + 0.0119 \, \bvel. \label{eq:lqr_optimized}
\end{align}
For each experimental setup we design the \texttt{non-switching} and \texttt{switching} controllers from Section~\ref{sec:controldesign} using the pendulum model introduced in~\cite{Nauta2025} and knowledge of the WH constraint experienced by the controller execution.

Our comparison is based on three performance metrics, one for each of the states of the dynamical system.
We denote with $\bar{t}$ the iteration in which the initial swing up phase completes.
The first and most important performance metric is the percentage of time that the pendulum spends around the top position $\percupright$.
As we discount the swing up phase, we can define this as the number of iterations in which $|\theta(t)| < 0.2\,$radians, roughly corresponding to an angle of $11\,$degrees, divided by the total number of iterations as
\begin{equation*}
	\percupright = \sfrac{1}{(N-\bar{t})}\sum^{N}_{i=\bar{t}} \big[|\pang(i)|<0.2\big],
\end{equation*}
where $[\cdot]$ represents the Iverson bracket.
The values of $\percupright$ are obviously bounded between $0$ and $1$, with higher values meaning better closed-loop behavior.

We also look at the mean value of the squared angular velocity $\errsqupvel$ and of the squared base velocity $\errsqubvel$, as these indicate that the pendulum and the base have on average moved more, which is a clear sign of worse closed-loop behavior.
\begin{equation*}
	\errsqupvel = \sfrac{1}{(N-\bar{t})}\sum^{N}_{i=\bar{t}} {\pvel(i)^2}, \quad
	\errsqubvel = \sfrac{1}{(N-\bar{t})}\sum^{N}_{i=\bar{t}} {\bvel(i)^2}
\end{equation*}
For those metrics, lower values indicate better closed-loop behavior.

As the performance channel, we use $w$ to model external disturbances, that we assume to influence $\theta$ and $\dot{\theta}$ similarly, resulting in $B^w = [1 \,\,\, 1 \,\,\, 0]^\top$.
With $z$, we choose which quantities relate to what we consider a good behavior of the pendulum, that is selecting $1$ as weight for the pendulum angle, $0.25$ as weight for the pendulum angular velocity, and $0.1$ as weight for the base velocity.
This indicates that we weigh more heavily the variations of the pendulum angle compared to the angular and base velocity (which we consider the least important) and results in $C = [1 \,\,\, 0.25 \,\,\, 0.1]^\top$, $D = 0$, and $D^w = 0$.
Those matrices can be tuned to the user's liking in case a different behavior is desired.
For example, by $D \neq 0$ one can also include the size of control inputs in the performance measure.

\subsection{Kill and Zero, \anymiss{7}{10}}
\label{sec:exp:kz}
In this first set of experiments we design a \texttt{non-switching} and a \texttt{switching} controller following the method presented in Section~\ref{sec:controldesign}.
The \texttt{non-switching} controller is
\begin{align}\begin{split}
	\text{\texttt{non-switching}}\!: u_\mathrm{c} ={}& 0.5492 \, {\pang} + 0.0541 \, \pvel \\
	&+ 0.0181 \, \bvel -1.0812 \, u,
	\label{eq:static_7_10_killzero}
\end{split}\end{align}
while the \texttt{switching} controller is composed of $36$ different alternatives, that depend on the graph and the computational state $(\mu_i)_{i \in \{ t-s+1, \dots, t-1 \}}$ experienced at the current iteration (and is not reported here due to space limitations).

Table~\ref{tab:killzero_7_10_p05} and Table~\ref{tab:killzero_7_10_p10} show the performance of the different controllers in the experiments conducted under the \anymiss{7}{10} constraint using the \emph{Kill and Zero} strategy.
The values reported in the tables represent the average performance across these runs, with the deadline miss probability $\pmiss$ set to $0.5$ and $1.0$, respectively.

As the execution conditions become more extreme, i.e., with higher $\pmiss$, we observe a degradation in the performance of all controllers, though to varying extents.
Notably, the \texttt{lqr} controller, which is designed without any robustness-to-deadline-misses considerations, performs significantly worse under these harsher conditions.
At $\pmiss = 1.0$, the angular velocity and base velocity errors (quantified respectively by $\errsqupvel$ and $\errsqubvel$) for the \texttt{lqr} controller increase dramatically, indicating near-instability conditions.
In contrast, the \texttt{original} controller maintains a moderate degradation in performance, suggesting some inherent robustness.
Note that for both \texttt{lqr} and \texttt{original} controllers, stability could not be guaranteed solving Optimization Problem~\ref{optprob:analysis}.
The controllers designed to handle deadline misses, both \texttt{non-switching} (numerically obtained $\ell_2$-performance $\gamma = 17.82$) and \texttt{switching} ($\gamma = 17.34$) controller demonstrate improved performance over both the \texttt{original} and \texttt{lqr} controllers, especially looking at $\errsqupvel$ and $\errsqubvel$ and in particular with higher probability of missing deadlines.
The outcome of this experimental setup validates our design approach.
By explicitly incorporating knowledge of the WH constraint into the controller synthesis, the designed controller can better manage and counteract frequent misses.

\begin{table}[t]
	\centering
	\caption{Kill and Zero, \anymiss{7}{10}, $\pmiss=0.5$.\\Averages over $100$ experiments.}
	\label{tab:killzero_7_10_p05} \vspace{-2mm}
	\rowcolors{2}{black!10}{white}
	\begin{tabular}{l|cc|cc}
		\hlinewd{1pt} \rowcolor{black!30}
		controller & avg $\percupright$ & std $\percupright$ & avg $\errsqupvel$ & avg $\errsqubvel$ \\
		\hlinewd{1pt}
		\texttt{original}~\eqref{eq:original}   & 0.97 & 0.16 & 1.35 & 3.57 \\ \hline
		\texttt{lqr}~\eqref{eq:lqr_optimized}        & 0.98 & 0.13 & 1.33 & 4.25 \\ \hline
		\texttt{non-switching}~\eqref{eq:static_7_10_killzero}     & 0.96 & 0.18 & 1.14 & 3.79 \\ \hline
		\texttt{switching}  & 0.96 & 0.17 & 1.52 & 4.74 \\ \hline
	\end{tabular}
	\vspace{4mm}
	\caption{Kill and Zero, \anymiss{7}{10}, $\pmiss=1.0$.\\Averages over $100$ experiments.}
	\label{tab:killzero_7_10_p10} \vspace{-2mm}
	\rowcolors{2}{black!10}{white}
	\begin{tabular}{l|cc|cc}
		\hlinewd{1pt} \rowcolor{black!30}
		controller & avg $\percupright$ & std $\percupright$ & avg $\errsqupvel$ & avg $\errsqubvel$ \\
		\hlinewd{1pt}
		\texttt{original}~\eqref{eq:original}   & 0.96 & 0.17 & \phantom{0}1.85 & \phantom{0}4.23 \\ \hline
		\texttt{lqr}~\eqref{eq:lqr_optimized}        & 0.18 & 0.18 &           25.21 &           20.21 \\ \hline
		\texttt{non-switching}~\eqref{eq:static_7_10_killzero}     & 0.95 & 0.18 & \phantom{0}1.69 & \phantom{0}4.11 \\ \hline
		\texttt{switching}  & 0.97 & 0.16 & \phantom{0}1.50 & \phantom{0}3.37 \\ \hline
	\end{tabular}
\end{table}

\subsection{Kill and Hold, \anymiss{3}{5}}
\label{sec:exp:kh}
In this second set of experiments we design a \texttt{non-switching} and a \texttt{switching} controller for this constraint and deadline management strategy. The non-switching controller has the following form
\begin{align}\begin{split}
	\text{\texttt{non-switching}}\!: u_\mathrm{c} ={}& 0.1429 \, {\pang} + 0.0191 \, \pvel \\
	&+ 0.0068 \, \bvel -0.3682 \, u,
	\label{eq:static_3_5_killhold}
\end{split}\end{align}
while the \texttt{switching} controller has $4$ different variants according to the graph for the WH constraint, that respectively are executed when the computational state $(\mu_i)_{i \in \{ t-s+1, \dots, t-1 \}}$ ends with $1$, with $10$, with $100$, or with $1000$, covering all the possible alternatives:
\begin{equation}\begin{split}
	\text{\texttt{switching}}\!: u_\mathrm{c} =&{} \\
	|\,(\makebox[\widthof{$0$}]{*}\makebox[\widthof{$0$}]{*}\makebox[\widthof{$0$}]{*}1)\,\, 0.1561 \, {\pang} &+ 0.0217 \, \pvel + 0.0078 \, \bvel -0.4114 \, u \\
	|\,(\makebox[\widthof{$0$}]{*}\makebox[\widthof{$0$}]{*}10)\,\, 0.1901 \, {\pang} &+ 0.0279 \, \pvel + 0.0103 \, \bvel -0.5186 \, u \\
	|\,(\makebox[\widthof{$0$}]{*}100)\,\, 0.2421 \, {\pang} &+ 0.0375 \, \pvel + 0.0142 \, \bvel -0.6879 \, u \\
	|\,(1000)\,\, 0.3569 \, {\pang} &+ 0.0597 \, \pvel + 0.0230 \, \bvel -1.0787 \, u.
	\label{eq:switching_3_5_killhold}
\end{split}\end{equation}
Note that a higher number of deadline misses makes the controller more aggressive in its own control signal calculation, i.e., the coefficients in the controller are all larger in absolute value compared to the alternative where fewer deadline misses have been experienced.

Table~\ref{tab:killhold_3_5_p05} and Table~\ref{tab:killhold_3_5_p10} present the results obtained under the \anymiss{3}{5} constraint using the \emph{Kill and Hold} strategy.
Compared to the \emph{Kill and Zero} case, this setting represents a more challenging control scenario, as confirmed by stability analyses presented in related works such as~\cite{Vreman2022b}.
The \emph{Hold} choice imposes tighter requirements on controller robustness, which are reflected in the performance results across the different control strategies.
The experimental results highlight the increasing difficulty of maintaining good closed-loop behavior using \emph{Hold}.
In both setups, the \texttt{lqr} controller exhibits a dramatic degradation in performance, with the values of $\errsqupvel$ and $\errsqubvel$ growing by more than two orders of magnitude, indicating loss of control, confirmed by the average percentage of time spent in the upright position being very low.
The corresponding LMI problem also reported infeasibility, indicating that the system might be unstable.
Interestingly enough, the case with $\pmiss=0.5$ is worse than when $\pmiss=1.0$, with the average value of $\percupright$ being respectively just above $0.2$ and just above $0.5$.
The \texttt{original} ($\gamma = 298.59$) controller also fails to maintain stability for some periods of time (with an average $\percupright$ of $0.82$).
In contrast, the \texttt{non-switching} ($\gamma = 7.84$) and \texttt{switching} ($\gamma = 6.21$) controllers presented in this paper and designed with awareness of the WH constraint show remarkable robustness.
The \texttt{switching} controller in particular achieves the best results in both angular and base velocity errors.
These results confirm that explicitly incorporating the constraint into the controller synthesis yields significant performance benefits in more complex scenarios.

\begin{table}[t] \centering
	\caption{Kill and Hold, \anymiss{3}{5}, $\pmiss=0.5$.\\Averages over $100$ experiments.}
	\label{tab:killhold_3_5_p05} \vspace{-2mm}
	\rowcolors{2}{black!10}{white}
	\begin{tabular}{l|cc|cc}
		\hlinewd{1pt} \rowcolor{black!30}
		controller & avg $\percupright$ & std $\percupright$ & avg $\errsqupvel$ & avg $\errsqubvel$ \\
		\hlinewd{1pt}
		\texttt{original}~\eqref{eq:original} & 0.91 & 0.12 & \phantom{0}77.25 & \phantom{0}41.82 \\ \hline
		\texttt{lqr}~\eqref{eq:lqr_optimized} & 0.23 & 0.04 & 663.85 & 340.48 \\ \hline
		\texttt{non-switching}~\eqref{eq:static_3_5_killhold} & 0.99 & 0.07 & \phantom{00}0.68 & \phantom{00}5.28 \\ \hline
		\texttt{switching}~\eqref{eq:switching_3_5_killhold} & 1.00 & 0.04 & \phantom{00}0.64 & \phantom{00}3.30 \\ \hline
	\end{tabular}
	\vspace{4mm}
	\caption{Kill and Hold, \anymiss{3}{5}, $\pmiss=1.0$.\\Averages over $100$ experiments.}
	\label{tab:killhold_3_5_p10} \vspace{-2mm}
	\rowcolors{2}{black!10}{white}
	\begin{tabular}{l|cc|cc}
		\hlinewd{1pt} \rowcolor{black!30}
		controller & avg $\percupright$ & std $\percupright$ & avg $\errsqupvel$ & avg $\errsqubvel$ \\
		\hlinewd{1pt}
		\texttt{original}~\eqref{eq:original}   & 0.82 & 0.14 &           144.31 & \phantom{0}86.09 \\ \hline
		\texttt{lqr}~\eqref{eq:lqr_optimized}        & 0.52 & 0.06 &           371.88 &           232.25 \\ \hline
		\texttt{non-switching}~\eqref{eq:static_3_5_killhold}     & 1.00 & 0.00 & \phantom{00}0.83 & \phantom{00}5.25 \\ \hline
		\texttt{switching}~\eqref{eq:switching_3_5_killhold}  & 1.00 & 0.00 & \phantom{00}0.72 & \phantom{00}3.53 \\ \hline
	\end{tabular}
\end{table}

\subsection{Skip and Zero, \anymiss{3}{5}}
\label{sec:exp:sz}
In this set of experiments we again use the \anymiss{3}{5} constraint and $\pmiss \in \{ 0.5, 1.0 \}$.
Solving the controller synthesis problem\footnote{We found experimentally that \emph{Skip} and \emph{Zero} requires a higher solver tolerance to perform well, as this strategy combination is challenging to control.} gives us the following two expressions, respectively for \texttt{non-switching} and \texttt{switching} controllers.
\begin{equation}
	\begin{aligned} \label{eq:static_3_5_skipzero}
		\text{\texttt{non-switching}}\!: u_\mathrm{c} ={}& 0.5065 \, {\pang} + 0.0577 \, \pvel \\
		&+ 0.0202 \, \bvel -1.0996 \, u
	\end{aligned}
\end{equation}
\vspace{-0.6em}
\begin{equation}
	\begin{aligned} \label{eq:switching_3_5_skipzero}
		\text{\texttt{switching}}\!: u_\mathrm{c} ={}& \\
		|\,(\makebox[\widthof{$0$}]{*}\makebox[\widthof{$0$}]{*}\makebox[\widthof{$0$}]{*}1)\,\, 0.4769 \, {\pang} &+ 0.0599 \, \pvel + 0.0217 \, \bvel -1.1294 \, u\\
		|\,(\makebox[\widthof{$0$}]{*}\makebox[\widthof{$0$}]{*}01)\,\, 0.4899 \, {\pang} &+ 0.0622 \, \pvel + 0.0225 \, \bvel -1.1716 \, u \\
		|\,(\makebox[\widthof{$0$}]{*}001)\,\, 0.4784 \, {\pang} &+ 0.0613 \, \pvel + 0.0222 \, \bvel -1.1528 \, u \\
		|\,(0001)\,\, 0.4365 \, {\pang} &+ 0.0545 \, \pvel + 0.0196 \, \bvel -1.0287 \, u
	\end{aligned}
\end{equation}

\begin{table}[t] \centering
	\caption{Skip and Zero, \anymiss{3}{5}, $\pmiss=0.5$.\\Averages over $100$ experiments.}
	\label{tab:skipzero_3_5_p05} \vspace{-2mm}
	\rowcolors{2}{black!10}{white}
	\begin{tabular}{l|cc|cc}
		\hlinewd{1pt} \rowcolor{black!30}
		controller & avg $\percupright$ & std $\percupright$ & avg $\errsqupvel$ & avg $\errsqubvel$ \\
		\hlinewd{1pt}
		\texttt{original}~\eqref{eq:original} & 0.98 & 0.11\phantom{0} & \phantom{0}1.83 & \phantom{0}4.10 \\ \hline
		\texttt{lqr}~\eqref{eq:lqr_optimized} & 0.96 & 0.17\phantom{0} & \phantom{0}3.45 & \phantom{0}6.18 \\ \hline
		\texttt{non-switching}~\eqref{eq:static_3_5_skipzero} & 0.96 & 0.12\phantom{0} & 10.19 & 16.72 \\ \hline
		\texttt{switching}~\eqref{eq:switching_3_5_skipzero} & 0.99 & 0.002 & \phantom{0}5.75 & 10.44 \\ \hline
	\end{tabular}
	\vspace{4mm}
	\caption{Skip and Zero, \anymiss{3}{5}, $\pmiss=1.0$.\\Averages over $100$ experiments.}
	\label{tab:skipzero_3_5_p10} \vspace{-2mm}
	\rowcolors{2}{black!10}{white}
	\begin{tabular}{l|cc|cc}
		\hlinewd{1pt} \rowcolor{black!30}
		controller & avg $\percupright$ & std $\percupright$ & avg $\errsqupvel$ & avg $\errsqubvel$ \\
		\hlinewd{1pt}
		\texttt{original}~\eqref{eq:original} & 0.79 & 0.29 & 10.52 & 11.73 \\ \hline
		\texttt{lqr}~\eqref{eq:lqr_optimized} & 0.97 & 0.11 & \phantom{0}7.09 & 11.80 \\ \hline
		\texttt{non-switching}~\eqref{eq:static_3_5_skipzero} & 0.97 & 0.15 & \phantom{0}5.39 & 11.76 \\ \hline
		\texttt{switching}~\eqref{eq:switching_3_5_skipzero} & 1.00 & 0.00 & \phantom{0}1.95 & \phantom{0}6.53 \\ \hline
	\end{tabular}
\end{table}

Table~\ref{tab:skipzero_3_5_p05} and Table~\ref{tab:skipzero_3_5_p10} respectively show the results obtained with the \texttt{original} and \texttt{lqr} controllers and with the controllers designed as presented in Section~\ref{sec:controldesign}.
At a moderate miss probability ($\pmiss = 0.5$), the \texttt{original} ($\gamma = 9096.5$) and \texttt{lqr} ($\gamma = 96.69$) controllers perform relatively well in terms of average upright time $\percupright$, though the \texttt{lqr} controller already shows increased motion ($\errsqupvel, \errsqubvel$ are larger), indicating degraded control.
Interestingly, the \texttt{non-switching} ($\gamma = 13.65$) and \texttt{switching} ($\gamma = 10.46$) controllers designed with constraint awareness still stabilize the pendulum well, but perform slightly worse in this setting in terms of velocity metrics, particularly the \texttt{non-switching} controller, which exhibits the highest values of $\errsqupvel$ and $\errsqubvel$.
However, at the extreme case of $\pmiss = 1.0$, the situation reverses: the \texttt{non-switching} and \texttt{switching} controllers maintain very good posture and control (high $\percupright$ and low velocity errors), whereas the \texttt{original} controller's performance deteriorates substantially.
Note also that the \texttt{switching} controller exhibits the best overall performance in almost all metrics, even when compared to the $\pmiss = 0.5$ cases.
This suggests that although constraint-aware controllers may not always outperform traditional ones under milder conditions, they exhibit superior robustness and maintain stability under high fault frequencies.
Thus, this experiment highlights the effectiveness of informed synthesis under harsh real-time execution scenarios.
Also, contrary to the \emph{Kill} experiments presented in Sections~\ref{sec:exp:kz} and~\ref{sec:exp:kh}, the \texttt{lqr} controller performs better than the \texttt{original} controller, showing that it is better optimized for the \emph{Skip} case.

\subsection{Skip and Hold, \anymiss{3}{5}}
\label{sec:exp:sh}
In this set of experiments we evaluate the system under the \emph{Skip and Hold} execution semantics with a WH constraint of \anymiss{3}{5}.
The synthesized controllers are given in the following equations.
\begin{equation}
	\begin{aligned}
		\text{\texttt{non-switching}}\!: u_\mathrm{c} ={}& 0.0907 \, {\pang} + 0.0102 \, \pvel \\
		&+ 0.0033 \, \bvel -0.0978 \, u \label{eq:static_3_5_skiphold}
	\end{aligned}
\end{equation}
\vspace{-0.6em} %
\begin{equation}
	\begin{aligned} \label{eq:switching_3_5_skiphold}
		\text{\texttt{switching}}\!: u_\mathrm{c} ={}& \\
		|\,(\makebox[\widthof{$0$}]{*}\makebox[\widthof{$0$}]{*}\makebox[\widthof{$0$}]{*}1)\,\, 0.1013 \, {\pang} &+ 0.0121 \, \pvel + 0.0041 \, \bvel -0.1138 \, u \\
		|\,(\makebox[\widthof{$0$}]{*}\makebox[\widthof{$0$}]{*}01)\,\, 0.1139 \, {\pang} &+ 0.0137 \, \pvel + 0.0047 \, \bvel -0.2399 \, u \\
		|\,(\makebox[\widthof{$0$}]{*}001)\,\, 0.1215 \, {\pang} &+ 0.0148 \, \pvel + 0.0051 \, \bvel -0.3112 \, u \\
		|\,(0001)\,\, 0.1184 \, {\pang} &+ 0.0144 \, \pvel + 0.0049 \, \bvel -0.2813 \, u
	\end{aligned}
\end{equation}

Tables~\ref{tab:skiphold_3_5_p05} and~\ref{tab:skiphold_3_5_p10} present the empirical results under this setup for the miss probabilities $\pmiss = 0.5$ and $\pmiss = 1.0$, respectively.
The switching controller~\eqref{eq:switching_3_5_skiphold} adapts its feedback gains based on the recent history of execution success and failures.
It distinguishes between four scenarios based on the number and pattern of deadline misses in the last four activations.
This conditional design allows the controller to increase its aggressiveness as more deadline misses accumulate.
Both the \texttt{original} and \texttt{lqr} controllers show severe instability, with very little time spent in the upright position and angular and base velocity errors increasing by over two orders of magnitude, confirming their inability to handle such aggressive fault patterns.
As a matter of fact, the corresponding LMI problems also report infeasibility, indicating that the systems might be unstable.

In stark contrast, both the \texttt{non-switching} ($\gamma = 58.41$) and \texttt{switching} ($\gamma = 30.38$) controllers maintain very good performance, while also keeping angular and base velocity errors within acceptable bounds.
Notably, the controllers perform better in extreme situations, which are closer to the worst-case they are designed for.
Also, the switching controller consistently outperforms the non-switching one in terms of both percentage of time spent in the upright position ($\percupright$) and motion smoothness ($\errsqupvel$ and $\errsqubvel$), demonstrating the benefits of a dynamic adaptation strategy that reacts to fault history.

\begin{table}[t] \centering
	\caption{Skip and Hold, \anymiss{3}{5}, $\pmiss=0.5$.\\Averages over $100$ experiments.}
	\label{tab:skiphold_3_5_p05} \vspace{-2mm}
	\rowcolors{2}{black!10}{white}
	\begin{tabular}{l|cc|cc}
		\hlinewd{1pt} \rowcolor{black!30}
		controller & avg $\percupright$ & std $\percupright$ & avg $\errsqupvel$ & avg $\errsqubvel$ \\
		\hlinewd{1pt}
		\texttt{original}~\eqref{eq:original} & 0.16 & 0.01 & 721.41 & 395.94 \\ \hline
		\texttt{lqr}~\eqref{eq:lqr_optimized} & 0.14 & 0.01 & 735.99 & 411.24 \\ \hline
		\texttt{non-switching}~\eqref{eq:static_3_5_skiphold} & 0.95 & 0.19 & \phantom{00}1.94 & \phantom{0}16.01 \\ \hline
		\texttt{switching}~\eqref{eq:switching_3_5_skiphold} & 0.99 & 0.05 & \phantom{00}2.14 & \phantom{0}12.01 \\ \hline
	\end{tabular}
	\vspace{4mm}
	\caption{Skip and Hold, \anymiss{3}{5}, $\pmiss=1.0$.\\Averages over $100$ experiments.}
	\label{tab:skiphold_3_5_p10} \vspace{-2mm}
	\rowcolors{2}{black!10}{white}
	\begin{tabular}{l|cc|cc}
		\hlinewd{1pt} \rowcolor{black!30}
		controller & avg $\percupright$ & std $\percupright$ & avg $\errsqupvel$ & avg $\errsqubvel$ \\
		\hlinewd{1pt}
		\texttt{original}~\eqref{eq:original} & 0.13 & 0.01 & 633.45 & 376.00 \\ \hline
		\texttt{lqr}~\eqref{eq:lqr_optimized} & 0.13 & 0.01 & 629.58 & 390.53 \\ \hline
		\texttt{non-switching}~\eqref{eq:static_3_5_skiphold} & 0.99 & 0.01 & \phantom{00}2.42 & \phantom{0}17.29 \\ \hline
		\texttt{switching}~\eqref{eq:switching_3_5_skiphold} & 0.99 & 0.05 & \phantom{00}2.14 & \phantom{0}12.01 \\ \hline
	\end{tabular}
\end{table}

\section{Conclusion}
\label{sec:concl}

This paper introduces a controller synthesis framework that explicitly integrates weakly-hard constraints into the controller design stage, proactively accounting for structured deadline misses and overrun semantics to connect real-time scheduling models with control-theoretic synthesis.
Using a switched systems formulation with a graph abstraction, the approach leverages semidefinite programming to guarantee stability and $\ell_2$-performance under arbitrary weakly-hard constraints, and is validated through extensive experiments on a Furuta pendulum where constraint-aware controllers consistently outperform traditional designs under aggressive fault and overload conditions, maintaining near-ideal stabilization even with frequent misses.
This highlights that anticipating and exploiting timing uncertainty during synthesis is more effective than merely tolerating it, and suggesting a path toward co-design with scheduling policies that may intentionally induce deadline misses when beneficial for other tasks.

\section*{Acknowledgments}
\noindent Marc Seidel thanks the Graduate Academy of the SC SimTech for its support.
Funded by the Deutsche Forschungsgemeinschaft (DFG, German Research Foundation) under Germany's Excellence Strategy -- EXC 2075 -- 390740016.
\vspace{2mm}\\
\noindent Martina Maggio is a member of the ELLIIT Strategic Research Area at Lund University.
This work was partially supported by the Wallenberg AI, Autonomous Systems and Software Program (WASP) funded by the Knut and Alice Wallenberg Foundation via the NEST projects \emph{Intelligent Cloud Robotics for Real-Time Manipulation at Scale} (\url{https://wasp-sweden.org/nest-project-cloud-robotics/}) and \emph{DYNACON: DYNamic Attack detection and mitigation for seCure AutONomy} (\url{https://wasp-sweden.org/nest-project-dynacon/}).

\bibliographystyle{plain}
\bibliography{references}
\balance
~

\end{document}